\documentclass[]{imsart}

\RequirePackage[OT1]{fontenc}
\RequirePackage{amsthm,amsmath}
\RequirePackage[numbers]{natbib}
\RequirePackage[colorlinks,citecolor=blue,urlcolor=blue]{hyperref}
\usepackage{amsmath,amssymb,amsthm}
\usepackage[toc,page]{appendix}
\usepackage{color}
\usepackage{graphics}
\usepackage{graphicx}
\usepackage{enumitem}
\usepackage{float}
\usepackage{booktabs}
\usepackage{array}
 \usepackage{verbatim}
  \newcommand{\F}{\mbox{$\cal F$}}

  \def\expe{\mathbb E}

  \def \convdist {\stackrel{d}{\longrightarrow}}

  \def\convp{\stackrel{p}{\longrightarrow}}
  
  \def\var{\mathrm{Var}}
  \def\cov{\mathrm{Cov}}
  
  \def\R{\mathbb R}

  \def \d {\displaystyle}
\startlocaldefs
\numberwithin{equation}{section}
\theoremstyle{plain}
  \newtheorem{thm}{Theorem}[section]
  \newtheorem{lem}[thm]{Lemma}

  \newtheorem{defn}[thm]{Definition}

\endlocaldefs

\begin{document}
\begin{frontmatter}
\title{How to verify that a given process is a L\'evy-Driven Ornstein-Uhlenbeck Process}
\runtitle{How to verify that a given process is an Ornstein-Uhlenbeck Processes}

% Add authors here manually without \begin{aug}
\author{Ibrahim Abdelrazeq$^{*}$, Hardy Smith, Dinmukhammed Zhanbyrshy}

% Add affiliations as footnotes
\footnotetext[1]{Department of Mathematics and Statistics, Rhodes College, 2000 North Pkwy, Memphis, TN 38112, USA, Emails: abdelrazeqi@rhodes.edu}

%\begin{aug}
%\author{\fnms{Ibrahim} \snm{Abdelrazeq}
%\ead[label=e1]{ abdelrazeqi@rhodes.edu}}
%\and
%\author{\fnms{H. Smith} \snm{Smith}\thanksref{t1}\ead[label=e2]{smirh-25@rhodes.edu}}
%\and
%\author{\fnms{D. Zhanbyrshy \snm{Zhanbyrshy}\thanksref{t1}\ead[label=e3]{zhadi@rhodes.edu}}

%\thankstext{t1}{Research supported by grants
 %from the Natural Sciences and Engineering %Research
 %Council of
 %Canada.}
 %\thankstext{t2}{Research supported by a grant
 %from the Natural Sciences and Engineering %Research
 %Council of
 %Canada.}

%\runauthor{I.\ Abdelrazeq}

%\address{Department of Mathematics and Statistics\\
% Rhodes College\\
%2000 North Pkwy, Memphis, TN 38112, USA \\
%\printead{e1}
%\affiliation{Rhodes College}

%\end{aug}

\begin{abstract}
Assuming that a L\'evy-Driven Ornstein-Uhlenbeck (or CAR(1)) processes is observed at discrete times $0$, $h$, $2h$,$\cdots$ $[T/h]h$. We introduce a step-by-step methodological approach on how a person would verify the model assumptions.  The methodology involves estimating the model parameters and approximating the driving process. We demonstrate how to use the increments of the approximated driving process, along with the estimated parameters, to test the assumptions that the CAR(1) process is L\'evy-driven. We then show how to test the hypothesis that the CAR(1) process belongs to a specified class of L\'evy processes. The performance of the tests is illustrated through multiple simulations. Finally, we demonstrate how to apply the methodology step-by-step to a variety of economic and financial data examples.
\end{abstract}

%\begin{keyword}[class=MSC]
%\kwd[Primary ]{60K35}
%\kwd{60K35}
%\kwd[; secondary ]{60K35}
%\end{keyword}

\tableofcontents

 \tableofcontents

\end{frontmatter}

\section{Introduction}

The continuous time analogue of the AR(1) time series is the Ornstein-Uhlenbeck process (equivalently, CAR(1) or CARMA(1,0) process) $Y$, which is the unique stationary solution of the stochastic differential equation
\begin{equation}\label{eq1}
dY(t)=-aY(t)dt+\sigma dL(t),~a,\sigma>0.
\end{equation}
The driving process $L$ is generally assumed to be L\'{e}vy - i.e. $L$ has independent increments. In \cite{Barndoroffandshephard2001},  Barndorff-Nielsen and Shephard used this process in a stochastic volatility model, and it has subsequently received considerable attention in the literature. As Estate Khmaladze points out in \cite{Khmaladze2021}, scanning through the database MathSciNet with key-word ``Ornstein-Uhlenbeck process" returned 18 pages of results, about 360 titles, all dated after year 2000. When the same is searched today, the database returned 68 pages of results, about 1360 titles, all dated after year 2000.\\

Between 2014 and 2018, Abdelrazeq et al. developed theories and results indicating how to develop a model verification for L\'evy-driven CAR(1) process [\cite{Ibrahim}, \cite{AIK}, \cite{CJS2018}]. Using these results, we develop a step-by-step methodological approach for how these findings can be used to verify the process and test its driving process distribution. We then show how to test the hypothesis that the CAR(1) process belongs to a specified class of L\'evy processes. The performance of these test statistics under the null hypothesis is illustrated via simulation studies. To demonstrate the power of the test, we computed the power under one alternative case. We then illustrate how to apply the methodology step-by-step to many economic and financial data examples. \\

 We hope that this will encourage many authors to adopt this methodology to verify the model before applying it. The objective of this paper is to make these results, along with the accompanying R code, accessible to users from all fields who may want to fit and apply this model to their data.\\

%While the probabilistic properties of the CAR(1) model are now well understood,  until recently there has been relatively little development in statistical inference for such models. 

When an observed data is assumed to follow a CAR(1) process, and since CAR(1) is a continuous-time model, inference for the CAR(1) process is complicated by the fact that the process $Y$ is typically sampled only at discrete times., i.e., $Y$ is observed at discrete times $0$, $h$, $2h$,$\cdots$, $[T/h]h$. This is the situation we consider here. As a result, the driving process $L$ cannot be observed directly and inference must be conducted with noisy data. The driving process must be approximated from the observed process. The increments of the approximated driving process, along with the estimated parameters, can be used to test the assumptions that the CAR(1) process is L\'evy-driven, Then, if the hypothesis that the driving process is L\'evy is not rejected, the increment can be used to test whether the driving process belongs to a specified class of L\'evy processes. \\

We proceed as follows. In Section \ref{sec:section 2}, we formally introduce the model and the test statistic. In Section \ref{sec:Section 3}, we detail our methodology for testing various background processes (Brownian motion, Gamma, Inverse Gaussian, etc.). We refer to Sections 1 and 2 of our supplemental material for a detailed explanation of our methodology, as well as corresponding R code. In Sections \ref{sec:section 4} and \ref{sec:section 5}, we produce examples that apply our test to high frequency financial data: S\&P 500 trading data and Euro/USD currency exchange data, respectively. We refer to Section 3 of our supplemental material for the R code corresponding to the S\&P 500 example in Section \ref{sec:section 4}.

\section{Preliminaries}\label{sec:section 2}

\subsection{L\'{e}vy processes}
  Suppose we are given a stochastic base $(\Omega, \F, (\F_t)_{0\leq t \leq \infty},P)$, where $\F_0$ contains all the $P$-null sets of $\F$ and $(\F_t)$ is right-continuous.
\begin{defn}[L\'{e}vy process] A process $L=\{L(t),t \geq 0\}$ is called a L\'{e}vy process if it is $(\F_t)$-adapted (i.e., $L(t)\in \F_t$  $ \forall~ t \ge 0$) and
\begin {itemize}
\item $L(0)=0$ a.s.
\item $L$ has independent increments, i.e., $L(t)-L(s)$ is independent of $\F_s$, for any $0 \leq s<t < \infty$.
\item $L$ has stationary increments, i.e., $L(t+s)-L(s)$ has the same distribution as  $L(t)$, for any $s,t >0$.
\item $L$ is stochastically continuous, i.e. $\forall~ \epsilon > 0$ and $\forall~ 0 \leq s<t < \infty $,
$$\lim_{s \to t}P(|L(t)-L(s)|>\epsilon )=0 .$$
\item $L$ has c\`{a}dl\`{a}g (right continuous with left limits) sample paths.
\end{itemize}
\end{defn}
\begin{defn}\label{second_order_levy_3}[Second-Order L\'evy Process]
We define $L$ to be a second-order L\'evy process if $L$ is a L\'evy process and
$\expe\left[ L^2(1)\right]<\infty$.   If  $\mu=\expe \left[
L(1)\right]$ and $\eta^2=\var \left[L(1)\right] $, then by the
independence and stationarity of the increments of $L(t)$ we have
 \begin{eqnarray}\label{second order levy}
\expe\left[L(t)\right]&=& \mu t, \textbf{{ } { } } t\ge 0 ,\nonumber \\
\var\left(L(t)\right)&=& \eta^2 t, \textbf{{ } { } } t\ge 0.
\end{eqnarray}
\end{defn}

Examples of second-order L\'evy processes include Brownian motion with drift, the Poisson process, Inverse Gaussian process, Beta process, and the Gamma process, which is characterized by $$ L(1) \sim \Gamma (\alpha, \beta)= \Gamma \left(\frac{\mu^2}{\eta^2}, \frac{\eta^2}{\mu}\right).$$

%the density should be in R
%$$f(x)= \frac{1}{\beta^\alpha \Gamma(\alpha)} x^{\alpha-1} e^{-\frac{x}{\beta}}$$
%{\bf Inverse Gaussian process}.\\

%$$L(t)\equiv IG(t)~~\text{where} ~~ IG_1 \sim IG(\delta, \gamma),~~ \delta>0, \gamma \ge 0,$$ where %$IG(\alpha, \beta)$ is a Inverse gaussian distribution  (also known as the Wald distribution). \\\\

%\textcolor{red}{Throughout this paper we assume that $L(t)$ is a second-order L\'evy process.}
\subsection{L\'evy-driven CAR(1) models}
In what follows, we assume that the process $L$ is c\`{a}dl\`{a}g with stationary increments.

\begin{defn}[CAR(1) process]\label{defn of Yt} A CAR(1) process $Y=\{Y(t),t\ge 0\}$ driven by the process $L=\{L(t),t\ge0\}$ is defined to be the solution of the stochastic differential equation
\begin{equation}\label{Eq:CAR1}
dY(t) = -a Y(t) dt+\sigma dL(t),
\end{equation}
where $a, \sigma \in \R_+$ and $Y(0)$ is independent of $\{L(t),t
\ge 0\}$. We call the process $L$ the driving process, and if $L$ is a L\'evy process then $Y$ is called a L\'evy-driven CAR(1) process.
\end{defn}

\begin{lem}[\cite{Barndoroffandshephard2001}]\label{lemma:moments_of_Y} Let $Y$ be a strictly stationary CAR(1) process driven by a second-order L\'evy process $L$ such that
(\ref{second order levy}) holds. Then for $s\geq 0$,
\begin{eqnarray}\label{expe and var and cov of y}
\expe\left[Y(0)\right]&=&\frac{\mu \sigma}{a},~~~\gamma_Y(s)\equiv \cov (Y(0), Y(s))= \frac{ \sigma^2 \eta^2 }{2a} e^{-as}.
\end{eqnarray}
\end{lem}

In this study, we consider the L\'evy-driven CAR(1) process sampled discretely at intervals of length $1/M$ over a time interval $[0,N]$, hence we have 
$N\times M$ observations sampled at $1/M,2/M,\ldots,N$. The following equations and Lemma \ref{CLT for sample coveriance} have been introduced and proved in \cite{AIK}, and we are including them here to introduce the concepts and provide a smooth thinking process for users.\\

Let $Y$ be a L\'evy-driven CAR(1) process, denote
\begin{eqnarray}\label{delta1Ln}
\Delta_1 L_n \equiv L_n- L_{n-1}
&=&\frac{1}{\sigma}\left[Y(n)-Y(n-1)+ a\int_{n-1}^{n} Y(s)ds\right],
\end{eqnarray}
to be the increment of the (unobserved) L\'evy process $L$ over the unit interval $[n-1,n]$. This unobserved increment, using the trapezoidal approximation as in \cite{Brockwelletall2007}, can be approximated by
\begin{eqnarray}\label{Delta_1 hatL_n}
\Delta_1\widehat{L}^{(M)}_n&\equiv&\frac{a}{M\sigma}\sum_{i=(n-1)M+1}^{nM}Y_\frac{i}{M}+ \left(\frac{1}{\sigma}-\frac{a}{2M\sigma}\right)\left(Y_{n}-Y_{n-1}\right).
\end{eqnarray}

 Let $\overline{\Delta_1\widehat{L}^{(M)}}$, $\widehat{\eta}^2$ and $\widehat{\gamma_{\Delta_1\widehat{L}^{(M)}}}(k)$ be the sample mean, sample variance and  sample covariances at lag $k\ge1$, respectively, of the estimated increments $\Delta_1 \widehat{L}^{(M)}_n$, i.e.
\begin{equation}\label{eta_defn}\overline {\Delta_1 \widehat{L}^{(M)}}\equiv \frac{1}{N} \sum_{n=1}^{N}\Delta_1 \widehat{L}^{(M)}_n,~~\widehat{\eta}^2 \equiv \frac{1}{N}\sum_{n=1}^{N}\left(\Delta_1 \widehat{L}^{(M)}_n -\overline{\Delta_1\widehat{L}^{(M)}} \right)^2
\end{equation}
and
 \begin{equation}\label{sample-cov-defn}
\widehat{\gamma_{\Delta_1\widehat{L}^{(M)}}}(k)\equiv \frac{1}{N-k}\sum_{n=1}^{N-k} \left(\Delta_1\widehat{L}^{(M)}_{n+k}-\overline{\Delta_1\widehat{L}^{(M)}}\right)
\left(\Delta_1\widehat{L}^{(M)}_n-\overline{\Delta_1\widehat{L}^{(M)}}\right).\end{equation}

\begin{lem}\label{CLT for sample coveriance} Under the assumptions of Lemma \ref{lemma:moments_of_Y}, we have:
\begin{itemize}
\item[{\rm (i)}] $\widehat{\gamma_{\Delta_1\widehat{L}^{(M)}}}(k) \convp 0 ~~\text{as}~~N\wedge M \to \infty~~\forall~ k\ge 1$;
\item[{\rm (ii)}] $\sqrt{N}~\widehat{\gamma_{\Delta_1\widehat{L}^{(M)}}}(k) \convdist N(0,\eta^4) ~~\text{as}~~N \to \infty~\text{and}~N/M\to 0~~\forall~ k\ge 1.$\\
\item[{\rm (iii)}] $W_{\Delta_1\widehat{L}^{(M)}}(k) \equiv\sqrt{N}~\frac{\widehat{\gamma_{\Delta_1\widehat{L}^{(M)}}}(k)}{\widehat{\eta}^2} \convdist N(0,1) ~~\text{as}~~N \to \infty~\text{and}~N/M\to 0.$
\end{itemize}
\end{lem}

\subsection{Test Statistic}\label{section 2.3} 

We define $W_{\widehat{\Delta_1\widehat{L}^{(M)}}}(1) $ to be the test statistic $W_{\Delta_1\widehat{L}^{(M)}}(k)$ defined in \cite{Ibrahim} if we replace the parameter $a$ by an estimator  $\widehat{a}^{(M)}_N$ to be specified in the next section and we define $\widehat{\widehat{\eta}^2}$ to be ${\widehat{\eta}^2}$ if we replace $a$ by $\widehat{a}^{(M)}_N$.\\

Let $Y_t$ be a discretely sampled stochastic process.\ If $Y$ is a CAR(1) model driven by a process $L$, we can use the estimated increments to test $H_0$ that $L$ has uncorrelated increments, which will be true if $L$ is a L\'evy process. We reject $H_0$ for a large absolute value of the statistic $W_{\widehat{\Delta_1\widehat{L}^{(M)}}}(1)$,\\

where 
\begin{equation}\label{test stat}
    W_{\widehat{\Delta_1\widehat{L}^{(M)}}}(1) \equiv\sqrt{N}~\frac{\widehat{\gamma_{\Delta_1\widehat{L}^{(M)}}}(1)}{\widehat{\eta}^2}.
\end{equation}

Under $H_0$, for large $N, M$ and $\frac{N}{M}$ small we have
 \begin{eqnarray}\label{levelalpha}
\alpha\approx P\left(|W_{\widehat{\Delta_1\widehat{L}^{(M)}}}(1)|>z_{\alpha/2}\right).
\end{eqnarray}

%%%%%%%%%%%%%%%%%%%%%%%%%%%%%%%%%%%%%%%%%

\section{Methodology}\label{sec:Section 3}
In this section, we outline the step-by-step methodology for testing various background processes. The approach and results are clearly presented to enable users to apply these techniques to their own models and data. For further explanation of the methodology, along with the corresponding R code, refer to Sections 1 and 2 of the supplemental material.

\subsection{Step 1: Choosing the right estimator for $a$}\label{step 1}

We are concerned with the consistency of the test statistic $W_{\Delta_1\widehat{L}^{(M)}}(1)$,  defined when the parameter $a$ is replaced by an estimator..Therefore, one must first select the most accurate estimator for $a$.\\

\textbf{Least squares based estimator}\label{sec:estimation-a}\\

\noindent We need to consider an estimator for $a$ for the general second order driving process $L$ (i.e. $\mu \neq 0$).  The following estimator of $a$:
\begin{equation}\label{ahat}
\widehat{{a}}_N^{(M)}= \frac{\sum_{n=1}^{NM} \left(Y_{\frac{n-1}{M}}-Y_{\frac{n}{M}}\right)\left(Y_{\frac{n-1}{M}}-\overline{Y}\right)}{\frac{1}{M}\sum_{n=1}^{NM} \left(Y_{\frac{n-1}{M}}-\overline{Y}\right)^2}~~\text{where}~~\overline{Y}=\frac{1}{NM}\sum_{n=1}^{NM}Y_\frac{n}{M},
\end{equation}

 is based on the least squares approach, as defined and studied in \cite{Ibrahim}; hence, we will refer to it as the least squares-based estimator (LSB). This estimator is recommended when the driving process is an unspecified L\'evy process or when Brownian motion is expected to be involved in the driving process. Additionally, it should be used to test whether the driving process is Brownian motion or not. The explanation and R code for this estimator can be found in Section 1.4.1 of the supplemental material.\\

\textbf{Davis-McCormick based estimator}\\

Brockwell et al. 2007 \cite{Brockwelletall2007} introduced an alternative estimator for $a$:

\begin{equation}
\widehat{a}_N^{(M)} = \sup_{0 \leq n < [NM]} \frac{\log(Y_{\frac{n}{M}}) - \log(Y_{\frac{n + 1}{M}})}{\frac{1}{M}}
\end{equation}\\

This estimator is based on the highly efficient Davis-McCormick estimator, as described in \cite{DavisMcCormick}; hence, we refer to it as the Davis-McCormick-based estimator (DMB estimator). It is highly accurate compared to the LSB estimator; however, the DMB estimator does not take negative values for $Y$. Therefore, if $Y$ is negative at any point, the LSB estimator should be used instead. Additionally, the DMB estimator should not be used when Brownian motion is expected to be involved in the driving process. We have tested and recommend this estimator for Gamma, Inverse Gaussian, Beta, and any mixed combinations of such  L\'evy processes. The explanation and R code for this estimator can be found in Section 1.4.2 of the supplemental material.\\

Below, in Tables 1-3, we compare the accuracy of our two estimators in simulated cases where the driving process is not Brownian motion. For these comparisons, we fix  $\sigma = 1$, $\mu = 1$, $\eta = 1$, $N = 100$, $M=100$, and $\lambda = 1 $ in the relevant cases.\\

\textit{Gamma driven CAR(1) process}

\begin{table}[H]
\caption{We fix $\{\sigma = 1, \mu = 1, \eta = 1, N=100, M=100\}$}
\vspace{\baselineskip}
\begin{tabular}{|c|c|c|}
\hline
$a$ & LSB & DMB \\ \hline
0.3 & 0.2905178    & 0.3000090    \\ \hline
0.9 & 1.0248610    & 0.9000810    \\ \hline
5   & 5.3899950    & 5.0025020    \\ \hline
10  & 9.6480590    & 10.010010    \\ \hline
\end{tabular}
\end{table}

\textit{Inverse Gaussian driven CAR(1) process}

\begin{table}[H]
\caption{We fix $\{\sigma = 1, \mu = 1, \eta = 1, N=100, M=100\}$}
\vspace{\baselineskip}
\begin{tabular}{|c|c|c|}
\hline
$a$ & LSB & DMB \\ \hline
0.3 & 0.2882148    & 0.2998861    \\ \hline
0.9 & 0.7399785    & 0.8998923    \\ \hline
5   & 5.2126420    & 5.0017530    \\ \hline
10  & 9.3897690    & 10.009330    \\ \hline
\end{tabular}
\end{table}

\textit{Mixed Inverse Gaussian and Gamma driven CAR(1) process}

\begin{table}[H]
\caption{We fix $\{\sigma = 1, \mu = 1, \eta = 1, N=100, M=100\}$}
\vspace{\baselineskip}
\begin{tabular}{|c|c|c|}
\hline
$a$ & LSB & DMB \\ \hline
0.3 & 0.2879226    & 0.2999216    \\ \hline
0.9 & 0.8566651    & 0.8998544    \\ \hline
5   & 4.7511070    & 5.0017370    \\ \hline
10  & 9.4394340    & 10.009760    \\ \hline
\end{tabular}
\end{table}

For the simulated cases with Gamma, Inverse Gaussian, or Mixed Inverse Gaussian and Gamma-driven processes, we observe that the DMB estimator is more accurate in estimation for all values of $a$ than the LSB estimator. Although the DMB estimator is more precise than the LSB estimator in most simulated cases, both estimators are relatively accurate for each background process. Moreover, the LSB estimator should be used when the driving process is Brownian motion or any unspecified process.\\

\subsection{Step 2: Choosing N and M}\label{step 2} 

Before calculating the recovered increments, one should choose $N$ and $M$ such that $N/M$ is small, as discussed in  \cite{AIK}. $N$ can represent a day, a month, or a year, with 
$M$ being the number of observations in each period accordingly. Additionally, we have observed that for small values of $a$ and $N$, the test may encounter issues with empirical level (i.e., the approximated p-value) being close to the nominal level. Therefore, we recommend that $N$ be greater than $50$.

\subsection{Step 3: Recovering increments with estimated $a$}\label{sec:estimated-a-main-result}

To proceed, we consider the recovered increments using an estimator for  $a$:
\begin{eqnarray}\label{delta with a hat}
\widehat{\Delta_1 \widehat{L}^{(M)}_{n}}&\equiv&\frac{\widehat{a}^{(M)}_N}{M}\sum_{i=(n-1)M+1}^{nM}Y_\frac{i}{M}+ \left(1-\frac{\widehat{a}^{(M)}_N}{2M}\right)\left(Y_n-Y_{n-1}\right)\nonumber\\
\end{eqnarray}

See Section 1.7 of our supplemental material for the R code, along with a detailed explanation of how to calculate the recovered increments with the estimated $a$.

\subsection{Step 4: Calculating the test statistic}\label{step 4}

We calculate $W_{\widehat{{\Delta_1\widehat{L}^{(M)}}}}(1)$ as defined in Equation (\ref{test stat}). For the significance level $\alpha = 0.05$, we reject the null hypothesis of uncorrelated increments when  $|W_{\widehat{{\Delta_1\widehat{L}^{(M)}}}}(1)| > 1.96$. In this case, we reject the null hypothesis that the data is L\'evy-driven for every background process. If $|W_{\widehat{{\Delta_1\widehat{L}^{(M)}}}}(1)| < 1.96$, we fail to reject the null hypothesis, and it is possible that a L\'evy-driven CAR(1) process is a good model. In this case, we recommend proceeding to Step 5 (see subsection \ref{step 5}). Refer to Section 1.8 of the supplemental material for instructions on how to calculate the test statistic in R.\\

\textbf{Performance of the test for various background processes}:\\

Under simulated  L\'evy-driven CAR(1) processes, we now examine the performance of our verification process for various background processes (i.e., Brownian motion, Gamma, Inverse Gaussian, Beta, and mixed L\'evy processes). The tables for each background process display the approximated p-value, for different values of $N$, $M$, and estimated $a$. See Section 2.1 of the supplemental material for the performance of the various background processes, along with the corresponding R code.\\

Note that a good indication of a test performing well under the null hypothesis is that, for a significance level of 0.05 and 400 simulations (i.e., $R=400$ in the following tables), the empirical level (i.e. the approximated p-value) should be around $0.05\pm 0.02$, which holds true in most of our cases. It is acceptable to have a deviation of 1 in 20, due to random chance, as predicted by the binomial distribution.\\

\textit{\textbf{Brownian motion driven CAR(1) process}}\\

\begin{table}[H]
\caption{We fix $\{\sigma = 1, \mu = 1, R = 400\}$}
\vspace{\baselineskip}
\begin{tabular}{|c|c|c|c|c|}
\hline
\multicolumn{1}{|l|}{} & $N=50, M=100$ & $N=100, M=100$ & $N=100, M=300$ & $N=100, M=500$ \\ \hline
\multicolumn{1}{|l|}{} & $\hat{\alpha}_{\triangle_1 \hat{BM}^{(M)}}$ & $\hat{\alpha}_{\triangle_1 \hat{BM}^{(M)}}$ & $\hat{\alpha}_{\triangle_1 \hat{BM}^{(M)}}$ & $\hat{\alpha}_{\triangle_1 \hat{BM}^{(M)}}$ \\ \hline
$a=0.1$ &0.0225 &0.0350 &0.0325 &0.0475 \\ \hline
$a=0.3$ &0.0100 &0.0300 &0.0125 &0.0225 \\ \hline
$a=0.5$ &0.0075 &0.0050 &0.0050 &0.0125 \\ \hline
$a=0.9$ &0.0200 &0.0225 &0.0175 &0.0175 \\ \hline
$a=3$   &0.0475 &0.0600 &0.0550 &0.0275 \\ \hline
$a=5$   &0.0375 &0.0525 &0.0600 &0.0450 \\ \hline
$a=7$   &0.0400 &0.0375 &0.0375 &0.0550 \\ \hline
$a=10$  &0.0375 &0.0575 &0.0475 &0.0425 \\ \hline
\end{tabular}
\end{table}

For the Brownian motion-driven process, we observe the approximated p-values very close to $\alpha = 0.05$ for most combinations of $N$, $M$, and $a$ values. It appears that the ratio  $N/M$ has no significant effect. However, for $a=0.3$, and $a=0.5$, we observe that the approximated p-values are not very close to the nominal level. In this table, we use LSB estimator to estimate $a$, as recommended for Brownian motion and refer to \cite{AIK} for further information on the effect of the LSB estimator on the results.\\

\textit{\textbf{Gamma driven CAR(1) process}}\\

\begin{table}[H]
\caption{We fix $\{\sigma = 1, \mu = 1, \eta = 1, R = 400\}$}
\vspace{\baselineskip}
\begin{tabular}{|c|c|c|c|c|}
\hline
\multicolumn{1}{|l|}{} & $N=50, M=100$ & $N=100, M=100$ & $N=100, M=300$ & $N=100, M=500$ \\ \hline
\multicolumn{1}{|l|}{} & $\hat{\alpha}_{\triangle_1 \hat{G}^{(M)}}$ & $ \hat{\alpha}_{\triangle_1 \hat{G}^{(M)}}$ & $\hat{\alpha}_{\triangle_1 \hat{G}^{(M)}}$ & $\hat{\alpha}_{\triangle_1 \hat{G}^{(M)}}$ \\ \hline
$a=0.1$ &0.0375 & 0.0400 & 0.0475 & 0.0275 \\ \hline
$a=0.3$ &0.0225 & 0.0325 & 0.0350 & 0.0425 \\ \hline
$a=0.5$ &0.0325 & 0.0525 & 0.0375 & 0.0450 \\ \hline
$a=0.9$ &0.0325 & 0.0375 & 0.0475 & 0.0350 \\ \hline
$a=3$   &0.0525 & 0.0350 & 0.0575 & 0.0475 \\ \hline
$a=5$   &0.0325 & 0.0300 & 0.0475 & 0.0475 \\ \hline
$a=7$   &0.0350 & 0.0400 & 0.0475 & 0.0425 \\ \hline
$a=10$  &0.0300 & 0.0450 & 0.0375 & 0.0425 \\ \hline
\end{tabular}
\end{table}

\textit{\textbf{Inverse Gaussian driven CAR(1) process}}\\

\begin{table}[H]
\caption{We fix $\{\sigma = 1, \mu = 1, \eta = 1, R = 400\}$}
\vspace{\baselineskip}
\begin{tabular}{|c|c|c|c|c|}
\hline
\multicolumn{1}{|l|}{} & $N=50, M=100$ & $N=100, M=100$ & $N=100, M=300$ & $N=100, M=500$ \\ \hline
\multicolumn{1}{|l|}{} & $\hat{\alpha}_{\triangle_1 \hat{IG}^{(M)}}$ & $\hat{\alpha}_{\triangle_1 \hat{IG}^{(M)}}$ & $\hat{\alpha}_{\triangle_1 \hat{IG}^{(M)}}$ & $\hat{\alpha}_{\triangle_1 \hat{IG}^{(M)}}$ \\ \hline
$a=0.1$ &0.0375 &0.0275 &0.0350 &0.0525 \\ \hline
$a=0.3$ &0.0250 &0.0500 &0.0575 &0.0350 \\ \hline
$a=0.5$ &0.0350 &0.0400 &0.0425 &0.0275 \\ \hline
$a=0.9$ &0.0500 &0.0300 &0.0375 &0.0300 \\ \hline
$a=3$   &0.0325 &0.0325 &0.0225 &0.0275 \\ \hline
$a=5$   &0.0400 &0.0400 &0.0200 &0.0200 \\ \hline
$a=7$   &0.0275 &0.0275 &0.0425 &0.0150 \\ \hline
$a=10$  &0.0400 &0.0500 &0.0250 &0.0350 \\ \hline
\end{tabular}
\end{table}

For the Gamma and Inverse Gaussian driven process, we observe the approximated p-values very close to $\alpha = 0.05$ for every combination of $N$, $M$, and $a$ values, and it appears that the ratio $N/M$ has no significant effect. In this table, we use the DMB estimator to estimate $a$, as recommended for these two cases.\\

\pagebreak

\textit{\textbf{Mixed Inverse Gaussian and Gamma driven CAR(1) process}}\\

\begin{table}[H]
\caption{We fix $\{\sigma = 1, \mu = 1, \eta = 1, R = 400\}$}
\vspace{\baselineskip}
\begin{tabular}{|c|c|c|c|c|}
\hline
\multicolumn{1}{|l|}{} & $N=50, M=100$ & $N=100, M=100$ & $N=100, M=300$ & $N=100, M=500$ \\ \hline
\multicolumn{1}{|l|}{} & $\hat{\alpha}_{\triangle_1 \hat{IGG}^{(M)}}$ & $\hat{\alpha}_{\triangle_1 \hat{IGG}^{(M)}}$ & $\hat{\alpha}_{\triangle_1 \hat{IGG}^{(M)}}$ & $\hat{\alpha}_{\triangle_1 \hat{IGG}^{(M)}}$ \\ \hline
$a=0.3$ &0.0350 &0.0300 &0.0225 &0.0325 \\ \hline
$a=0.9$ &0.0375 &0.0375 &0.0300 &0.0525 \\ \hline
$a=5$   &0.0300 &0.0350 &0.0275 &0.0275 \\ \hline
$a=10$  &0.0200 &0.0425 &0.0350 &0.0300 \\ \hline
\end{tabular}
\end{table}

For the mixed Inverse Gaussian and Gamma driven process, Similarly, as above, we observe the approximated p-values very close to $\alpha = 0.05$ for every combination of $N$, $M$, and $a$ values, and it appears that the ratio $N/M$ has no significant effect. In this table, we use the DMB estimator to estimate $a$, as recommended for this case.\\

Our results emphasize the strength of the verification process, as under the null hypothesis, we fail to reject the null hypothesis of uncorrelated increments for many L\'evy-driven background processes, with approximated p-values very close to the nominal level, across varying combinations of $N$, $M$, and $a$ values. See Section 2 of the supplemental material for testing the performance of various background processes.\\

\subsection{Step 5: Testing for the driving process}\label{step 5}

If the null hypothesis that the driving process is L\'evy is not rejected, then the L\'evy-driven CAR(1) could be considered a good model. In this case, the next step is to test whether the driving process is Brownian motion. Another interesting test is to determine whether the driving process is some other specific process, such as Gamma, Inverse Gaussian, or any other unspecified process. We recommend the following two procedures, generally adapted from \cite{CJS2018} and \cite{Stuteetal}, and modified for our verification process, to achieve these objectives.\\

\textbf{Procedure 1:}\\

To test whether the driving process is Brownian motion (BM), we follow these steps.

\begin{enumerate}
    \item Recover the estimated increments, as shown in Equation (\ref{sec:estimated-a-main-result}), using the LSB estimator. 
    \item Generate one bootstrap sample from the calculated estimated increments obtained in Step 1.
    \item Compute the sample mean and standard deviation of the bootstrap sample obtained in Step 2.
        \item Conduct the Kolmogorov-Smirnov (KS) test for the normal distribution using the sample mean and standard deviation calculated in Step 3. 
\end{enumerate} 

This procedure should only be used if the process being tested is Brownian motion, as explained in \cite{BurkeGombay1998} and \cite{CJS2018}..\\

\textbf{Procedure 2:}\\

To test whether the driving process is something other than Brownian motion, such as Gamma, Inverse Gaussian, or any other unspecified process, as outlined in \cite{Stuteetal}, we follow these steps:\\

\begin{enumerate}
    \item Recover the estimated increments, as shown in Equation \ref{sec:estimated-a-main-result}, using the appropriate estimator (LSB or DMB).
    \item Assuming the recovered increments follow a particular distribution $F(.;\theta)$, estimate the parameters for this distribution from the recovered increments.
    \item Evaluate $Z_{i}$ - the CDF values of all recovered increments - using the estimated parameters from Step 2
    \item Rearrange $Z_{i}$ in ascending order
    \item Calculate the test statistic $D_N$ for the $N$ recovered increments $Z_{i}$, where where $i$ ranges from $1, ..., N$, and $D_N$ is defined as 
\[D_{N} = N^{1/2} \underset{1 \leq i \leq N}{\mathrm{max}} \left\{\frac{i}{N} - Z_{i:N}, Z_{i:N} - \frac{(i-1)}{N}\right\}\] 
    \item Generate 1000 random bootstrap samples from the assumed distribution $F(.;\theta)$, using the estimated parameters from Step 2. For each bootstrap sample:
    \begin{enumerate}[label=\alph*)]
        \item Estimate the bootstrap parameters as in Step 2.
        \item Evaluate $Z_{i}$ using the estimated bootstrap parameters.
        \item Rearrange $Z_{i}$ in ascending order.
        \item Calculate the test statistic $D_N$ using the formula from Step 5
    \end{enumerate}
    \item Compare the original test statistic from Step 5 to the 95th percentile critical value of the bootstrap distribution. If the original test statistic exceeds this value, reject the null hypothesis.
\end{enumerate}

Theoretically, this procedure can be used when testing for any specified process completely determined by the mean $\mu$ and variance $\eta^2$, as detailed in \cite{CJS2018}. However, we have only successfully applied it when the driving processes are Gamma, Inverse Gaussian, and Brownian motion. As mentioned earlier, we recommend using Procedure 1 when testing for Brownian motion, as it is computationally easier.\\

Under the null hypothesis $H_0$, we expect the estimated p-value for the test to be around the nominal level ($\alpha=0.05$). In Table 8 below, we use Procedure 1 to test whether the driving process is Brownian motion, with varying values of $N$, $M$, and $a$. The corresponding R code can be found in Section 2.2.1 of the supplemental material.\\

\textit{Brownian motion driven CAR(1) process using Procedure 1}:

\begin{table}[H]
\caption{We fix $\{\sigma = 1, \mu = 1, \eta = 1, R = 400\}$}
\vspace{\baselineskip}
\begin{tabular}{|c|c|c|c|c|}
\hline
\multicolumn{1}{|l|}{} & $N=50, M=100$ & $N=100, M=100$ & $N=100, M=300$ & $N=100, M=500$ \\ \hline
\multicolumn{1}{|l|}{} & $\hat{\alpha}_{\triangle_1 \hat{BM}^{(M)}}$ & $\hat{\alpha}_{\triangle_1 \hat{BM}^{(M)}}$ & $\hat{\alpha}_{\triangle_1 \hat{BM}^{(M)}}$ & $\hat{\alpha}_{\triangle_1 \hat{BM}^{(M)}}$ \\ \hline
$a=0.3$ &0.0525 & 0.0400 & 0.0225 & 0.0575 \\ \hline
$a=0.9$ &0.0350 & 0.0325 & 0.0400 & 0.0400 \\ \hline
$a=5$   &0.0475 & 0.0400 & 0.0325 & 0.0475 \\ \hline
$a=10$  &0.0450 & 0.0400 & 0.0250 & 0.0325 \\ \hline
\end{tabular}
\end{table}

When the driving process is Brownian motion, we observe p-values below or very near 0.05 for each combination of $N$, $M$, and $a$ values when using Procedure 1.\\

In Tables 9 and 10 below, we use Procedure 2 to test whether the driving process is a specific process, namely Gamma in Table 9 and Inverse Gaussian in Table 10, under the null hypothesis $H_0$. The corresponding R code can be found in Section 2.2.2 of the supplemental material.\\

\textit{Gamma driven CAR(1) process}:

\begin{table}[H]
\caption{We fix $\{\sigma = 1, \mu = 1, \eta = 1, R = 400\}$}
\vspace{\baselineskip}
\begin{tabular}{|c|c|c|c|c|}
\hline
\multicolumn{1}{|l|}{} & $N=50, M=100$ & $N=100, M=100$ & $N=100, M=300$ & $N=100, M=500$ \\ \hline
\multicolumn{1}{|l|}{} & $\hat{\alpha}_{\triangle_1 \hat{G}^{(M)}}$ & $\hat{\alpha}_{\triangle_1 \hat{G}^{(M)}}$ & $\hat{\alpha}_{\triangle_1 \hat{G}^{(M)}}$ & $\hat{\alpha}_{\triangle_1 \hat{G}^{(M)}}$ \\ \hline
$a=0.3$ &0.0375 & 0.0475 & 0.0300 & 0.0425 \\ \hline
$a=0.9$ &0.0525 & 0.0525 & 0.0725 & 0.0375 \\ \hline
$a=5$   &0.0475 & 0.0425 & 0.0475 & 0.0725 \\ \hline
$a=10$  &0.0600 & 0.0575 & 0.0550 & 0.0600 \\ \hline
\end{tabular}
\end{table}

\textit{Inverse Gaussian driven CAR(1) process}:

\begin{table}[H]
\caption{We fix $\{\sigma = 1, \mu = 1, \eta = 1, R = 400\}$}
\vspace{\baselineskip}
\begin{tabular}{|c|c|c|c|c|}
\hline
\multicolumn{1}{|l|}{} & $N=50, M=100$ & $N=100, M=100$ & $N=100, M=300$ & $N=100, M=500$ \\ \hline
\multicolumn{1}{|l|}{} & $\hat{\alpha}_{\triangle_1 \hat{IG}^{(M)}}$ & $\hat{\alpha}_{\triangle_1 \hat{IG}^{(M)}}$ & $\hat{\alpha}_{\triangle_1 \hat{IG}^{(M)}}$ & $\hat{\alpha}_{\triangle_1 \hat{IG}^{(M)}}$ \\ \hline
$a=0.3$ &0.0800 & 0.0725 & 0.0625 & 0.0500 \\ \hline
$a=0.9$ &0.0400 & 0.0500 & 0.0650 & 0.0425 \\ \hline
$a=5$   &0.0675 & 0.0700 & 0.0475 & 0.0350 \\ \hline
$a=10$  &0.0625 & 0.0500 & 0.0475 & 0.0775 \\ \hline
\end{tabular}
\end{table}

When the driving process is Gamma or Inverse Gaussian , we observe p-values below or very near 0.05 for each combination of $N$, $M$, and $a$ values.\\

\subsubsection{Power of the test}

When the process is not driven by Brownian motion, such as Gamma, Inverse Gaussian, or mixed combinations of these, we expect high rejection rate (say $\hat{\beta}_{\triangle_1 \hat{L}^{(M)}}$  for the test when testing for Brownian motion. In Tables 11-13 below, the results for the test power are presented when using Procedure 1.\\

\textit{Simulating Gamma driven CAR(1) process and testing for BM using Procedure 1}:

\begin{table}[H]
\caption{We fix $\{\sigma = 1, \mu = 1, \eta = 1, R = 400\}$}
\vspace{\baselineskip}
\begin{tabular}{|c|c|c|c|c|}
\hline
\multicolumn{1}{|l|}{} & $N=50, M=100$ & $N=100, M=100$ & $N=100, M=300$ & $N=100, M=500$ \\ \hline
\multicolumn{1}{|l|}{} & $\hat{\beta}_{\triangle_1 \hat{G}^{(M)}}$ & $\hat{\beta}_{\triangle_1 \hat{G}^{(M)}}$ & $\hat{\beta}_{\triangle_1 \hat{G}^{(M)}}$ & $\hat{\beta}_{\triangle_1 \hat{G}^{(M)}}$ \\ \hline
$a=0.3$ &0.4550 & 0.8350 & 0.8400 & 0.8550 \\ \hline
$a=0.9$ &0.4600 & 0.8450 & 0.9050 & 0.8725 \\ \hline
$a=5$   &0.4500 & 0.8900 & 0.9075 & 0.9000 \\ \hline
$a=10$  &0.5200 & 0.9225 & 0.8950 & 0.8750 \\ \hline
\end{tabular}
\end{table}

\textit{Simulating Inverse Gaussian driven CAR(1) process and testing for BM using Procedure 1}:

\begin{table}[H]
\caption{We fix $\{\sigma = 1, \mu = 1, \eta = 1, R = 400\}$}
\vspace{\baselineskip}
\begin{tabular}{|c|c|c|c|c|}
\hline
\multicolumn{1}{|l|}{} & $N=50, M=100$ & $N=100, M=100$ & $N=100, M=300$ & $N=100, M=500$ \\ \hline
\multicolumn{1}{|l|}{} & $\hat{\beta}_{\triangle_1 \hat{IG}^{(M)}}$ & $\hat{\beta}_{\triangle_1 \hat{IG}^{(M)}}$ & $\hat{\beta}_{\triangle_1 \hat{IG}^{(M)}}$ & $\hat{\beta}_{\triangle_1 \hat{IG}^{(M)}}$ \\ \hline
$a=0.3$ &0.6025 & 0.9425 & 0.9425 & 0.9475 \\ \hline
$a=0.9$ &0.6125 & 0.9700 & 0.9425 & 0.9575 \\ \hline
$a=5$   &0.6975 & 0.9425 & 0.9675 & 0.9650 \\ \hline
$a=10$  &0.6700 & 0.9675 & 0.9600 & 0.9750 \\ \hline
\end{tabular}
\end{table}

\textit{Simulating mixed Inverse Gaussian and Gamma driven CAR(1) process and testing for BM using Procedure 1}:

\begin{table}[H]
\caption{We fix $\{\sigma = 1, \mu = 1, \eta = 1, R = 400\}$}
\vspace{\baselineskip}
\begin{tabular}{|c|c|c|c|c|}
\hline
\multicolumn{1}{|l|}{} & $N=50, M=100$ & $N=100, M=100$ & $N=100, M=300$ & $N=100, M=500$ \\ \hline
\multicolumn{1}{|l|}{} & $\hat{\beta}_{\triangle_1 \hat{IGG}^{(M)}}$ & $\hat{\beta}_{\triangle_1 \hat{IGG}^{(M)}}$ & $\hat{\beta}_{\triangle_1 \hat{IGG}^{(M)}}$ & $\hat{\beta}_{\triangle_1 \hat{IGG}^{(M)}}$ \\ \hline
$a=0.3$ &0.5450 & 0.9000 & 0.9150 & 0.8975 \\ \hline
$a=0.9$ &0.5475 & 0.8875 & 0.8975 & 0.9175 \\ \hline
$a=5$   &0.6625 & 0.9450 & 0.9400 & 0.9250 \\ \hline
$a=10$  &0.5800 & 0.9525 & 0.9200 & 0.9350 \\ \hline
\end{tabular}
\end{table}

When the driving process is Gamma, Inverse Gaussian, or a mixed Inverse Gaussian and Gamma process, we observe a high rejection rate ($\hat{\beta}_{\triangle_1 \hat{L}^{(M)}}$) for each combination of $N$, $M$, and $a$ values.\\

To illustrate the effectiveness of the test using Procedure 2, we observe the rejection rate ($\hat{\beta}_{\triangle_1 \hat{L}^{(M)}}$) for testing Brownian motion in Tables 14-16 below, where Procedure 2 is applied.\\

\textit{Simulating Gamma driven CAR(1) process and testing for BM using Procedure 2}:

\begin{table}[H]
\caption{We fix $\{\sigma = 1, \mu = 1, \eta = 1, R = 400\}$}
\vspace{\baselineskip}
\begin{tabular}{|c|c|c|c|c|}
\hline
\multicolumn{1}{|l|}{} & $N=50, M=100$ & $N=100, M=100$ & $N=100, M=300$ & $N=100, M=500$ \\ \hline
\multicolumn{1}{|l|}{} & $\hat{\beta}_{\triangle_1 \hat{G}^{(M)}}$ & $\hat{\beta}_{\triangle_1 \hat{G}^{(M)}}$ & $\hat{\beta}_{\triangle_1 \hat{G}^{(M)}}$ & $\hat{\beta}_{\triangle_1 \hat{G}^{(M)}}$ \\ \hline
$a=0.3$ &0.9175 & 1.0000 & 1.0000 & 1.0000 \\ \hline
$a=0.9$ &0.9175 & 1.0000 & 0.9975 & 1.0000 \\ \hline
$a=5$   &0.9525 & 1.0000 & 1.0000 & 1.0000 \\ \hline
$a=10$  &0.9525 & 1.0000 & 1.0000 & 1.0000 \\ \hline
\end{tabular}
\end{table}

\textit{Simulating Inverse Gaussian driven CAR(1) process and testing for BM using Procedure 2}:

\begin{table}[H]
\caption{We fix $\{\sigma = 1, \mu = 1, \eta = 1, R = 400\}$}
\vspace{\baselineskip}
\begin{tabular}{|c|c|c|c|c|}
\hline
\multicolumn{1}{|l|}{} & $N=50, M=100$ & $N=100, M=100$ & $N=100, M=300$ & $N=100, M=500$ \\ \hline
\multicolumn{1}{|l|}{} & $\hat{\beta}_{\triangle_1 \hat{IG}^{(M)}}$ & $\hat{\beta}_{\triangle_1 \hat{IG}^{(M)}}$ & $\hat{\beta}_{\triangle_1 \hat{IG}^{(M)}}$ & $\hat{\beta}_{\triangle_1 \hat{IG}^{(M)}}$ \\ \hline
$a=0.3$ &0.9775 & 1.0000 & 1.0000 & 1.0000 \\ \hline
$a=0.9$ &0.9950 & 1.0000 & 1.0000 & 1.0000 \\ \hline
$a=5$   &0.9875 & 1.0000 & 1.0000 & 1.0000 \\ \hline
$a=10$  &0.9825 & 1.0000 & 1.0000 & 1.0000 \\ \hline
\end{tabular}
\end{table}

\textit{Simulating mixed Inverse Gaussian and Gamma driven CAR(1) process and testing for BM using Procedure 2}:

\begin{table}[H]
\caption{We fix $\{\sigma = 1, \mu = 1, \eta = 1, R = 400\}$}
\vspace{\baselineskip}
\begin{tabular}{|c|c|c|c|c|}
\hline
\multicolumn{1}{|l|}{} & $N=50, M=100$ & $N=100, M=100$ & $N=100, M=300$ & $N=100, M=500$ \\ \hline
\multicolumn{1}{|l|}{} & $\hat{\beta}_{\triangle_1 \hat{IGG}^{(M)}}$ & $\hat{\beta}_{\triangle_1 \hat{IGG}^{(M)}}$ & $\hat{\beta}_{\triangle_1 \hat{IGG}^{(M)}}$ & $\hat{\beta}_{\triangle_1 \hat{IGG}^{(M)}}$ \\ \hline
$a=0.3$ &0.9650 & 1.0000 & 1.0000 & 1.0000 \\ \hline
$a=0.9$ &0.9850 & 1.0000 & 1.0000 & 1.0000 \\ \hline
$a=5$   &0.9600 & 1.0000 & 1.0000 & 1.0000 \\ \hline
$a=10$  &0.9725 & 1.0000 & 1.0000 & 1.0000 \\ \hline
\end{tabular}
\end{table}

When the driving process is Gamma, Inverse Gaussian, or a mixed Inverse Gaussian and Gamma process, we observe a high rejection rate ($\hat{\beta}_{\triangle_1 \hat{L}^{(M)}}$) for each combination of $N$, $M$, and $a$ values.\\

As a general observation, we find that both Procedure 1 and Procedure 2 exhibit very good power in all cases, with Procedure 2 having an advantage over Procedure 1. However, Procedure 1 is computationally easier, particularly when testing for Brownian motion.\\

Although the tables above show the power results for only one case-when the driving process is Brownian motion and we use Procedure 1 and Procedure 2-we have tested many other combinations of processes using Procedure 2. For example, when the driving process is Gamma and we test for Inverse Gaussian, and vice versa, the rejection rates are similar to or greater than those observed in the tables above.\\

We observe results that align with our expectations both under the null hypothesis ($H_0$) and for the power of the test. For various combinations of $N$, $M$ and $a$ values, the estimated p-values are around nominal level ($\alpha=0.05$) under $H_0$. For the test power, we observe high rejection rate  when testing for a range of different processes. The corresponding R code can be found in Section 2.2 of the supplemental material.\\

\section{Application to S\&P 500 Stocks Dynamic Spread}\label{sec:section 4}

\subsection{Data and model}

We apply our verification process to minute-by-minute stock prices of the S\&P 500 index constituents from December 2014 to December 2015. The R code for this application is provided in Section 3 of the supplemental material. The S\&P 500 is a major stock market index that tracks the performance of 500 of the largest publicly traded companies in the United States, serving as the benchmark index for large U.S. corporations. For each S\&P 500 constituent on any given day, our data matrix includes the date (represented as an integer), the time (minute-by-minute on a 24-hour clock), and the open, high, low, and close prices for each minute  \cite{QuantQuote2016}.\\

To implement our verification process, we define the price spread between two stocks, A and B, with prices  $S_A(t)$ and  $S_B(t)$ as follows:
 \begin{equation}
    Y_t = \ln\left(\frac{S_A(t)}{S_A(0)}\right) - \ln\left(\frac{S_B(t)}{S_B(0)}\right).
\end{equation}

Spread measures the relative change in the prices of two stocks over time. The goal is to evaluate whether these spread dynamics can be modeled using a L\'evy-driven CAR(1) process, Equation (\ref{Eq:CAR1}), as outlined in our methodology. In our example, We calculate the spread dynamics for several large-cap S\&P 500 constituent pairs within their corresponding Global Industry Classification Standard (GICS) sector (GICS) sectors.\\

The model for the spread dynamics of the pair Abbott Laboratories (ABT) and Danaher Corporation (DHR) is shown in Figure 1:

\begin{figure}[H]
  \centering
  \includegraphics[width=0.64\textwidth]{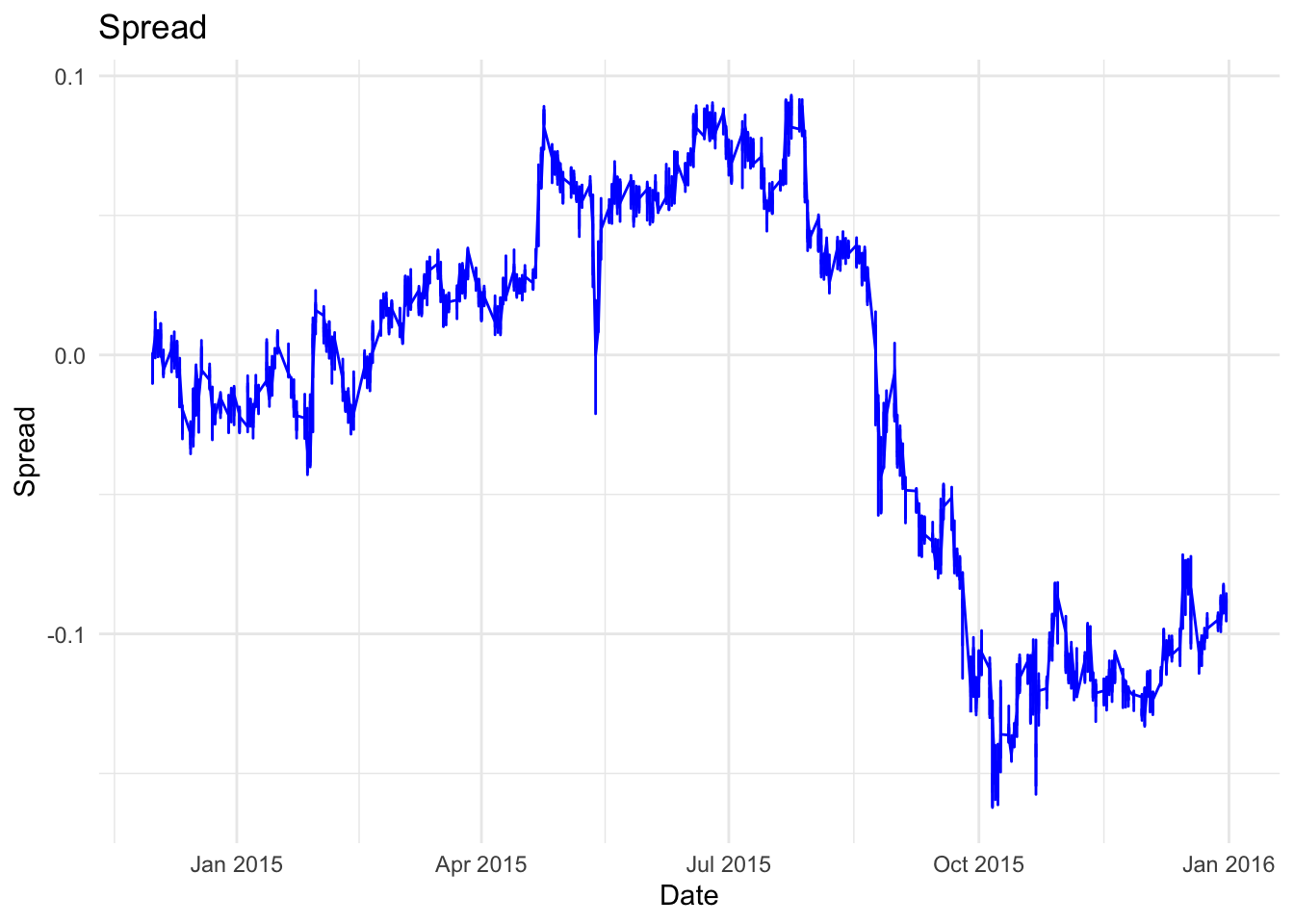}
  \caption{\label{Spread1} ABT and DHR spread dynamic}
\end{figure}

The model for the spread dynamics of the pair Amgen and Pfizer is shown in Figure 2:

\begin{figure}[H]
  \centering
  \includegraphics[width=0.64\textwidth]{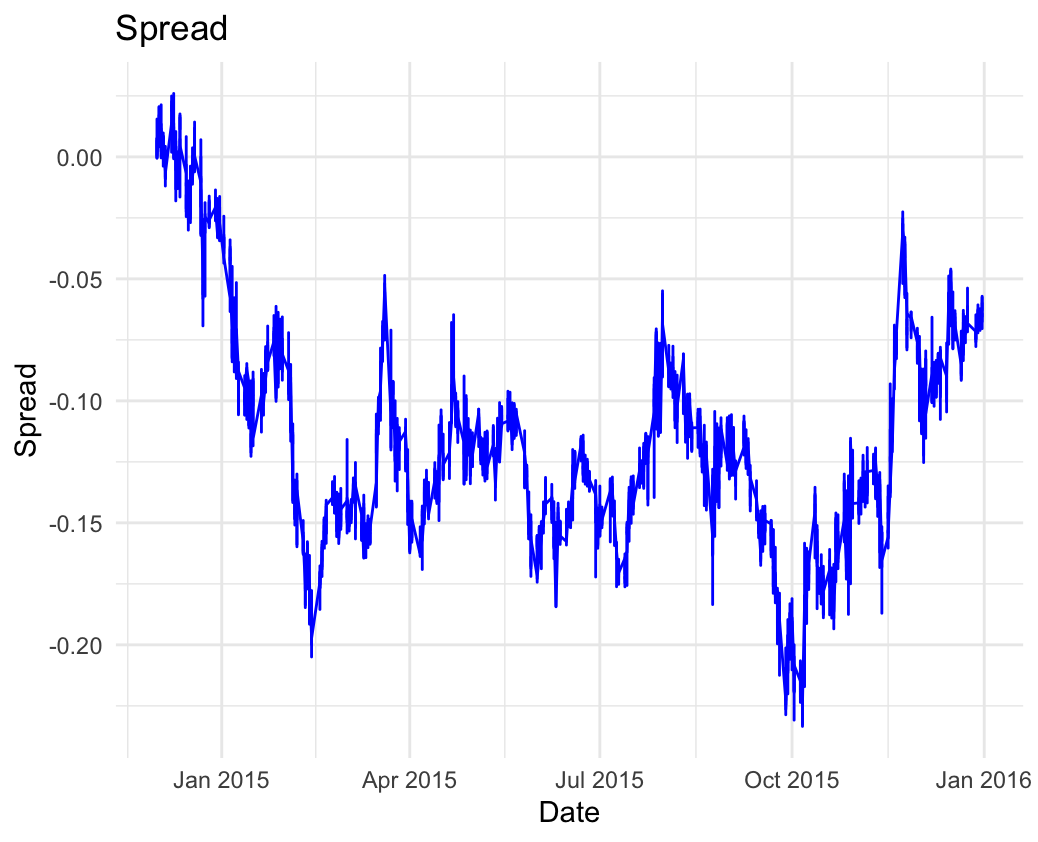}
  \caption{\label{Spread2} Amgen and Pfizer spread dynamic}
\end{figure}

\subsection{Verification}

We follow the methodology detailed in Section 3. In Step 1 (see Subsection 3.1), we select our estimator for $a$. Since the price spread can take on negative values, we use the LSB estimator to estimate $a$. In Step 2 (see Subsection 3.2), we determine the values for $N$ and $M$. The time period is provided by the data, specifically the observed spread dynamics from December 2014 to December 2015. Therefore, we set $M=500$ so that the spread dynamics are sampled at intervals of $1/M, 2/M, ..., N$. In Step 3 (see Subsection 3.3), we recover the spread dynamics using our chosen $M$ and estimated $a$. In Step 4 (see Subsection 3.4), we calculate the test statistic and test the correlation of the recovered increments. The application of this methodology to our data is presented in Section 3 of the supplemental material.\\

We model the spread of several combinations of large-cap stocks from corresponding Global Industry Classification Standard (GICS) sectors to verify that they are L\'evy-driven, with test statistics falling below the critical value. $z_{1-\alpha/2} = 1.96$.\\

We observe varying and sometimes inconclusive results from our test. For example, when measuring the price spread between Abbott Laboratories (ABT) and Danaher Corporation (DHR), we fail to reject the null hypothesis of uncorrelated increments, with a test statistic of 1.05. Similarly, the spread between Apple (AAPL) and Google (GOOG) also fails to reject the null hypothesis, yielding a test statistic of 1.85. This suggests that a L\'evy-driven Ornstein-Uhlenbeck (OU) model may be a suitable model for these spreads. In Figures 3 and 4 below, we present the recovered increments for the ABT-DHR spread both as a time series and as residuals, respectively.\\

\begin{figure}[H]
  \centering
  \begin{minipage}[b]{0.45\textwidth}
    \centering
    \includegraphics[width=\textwidth]{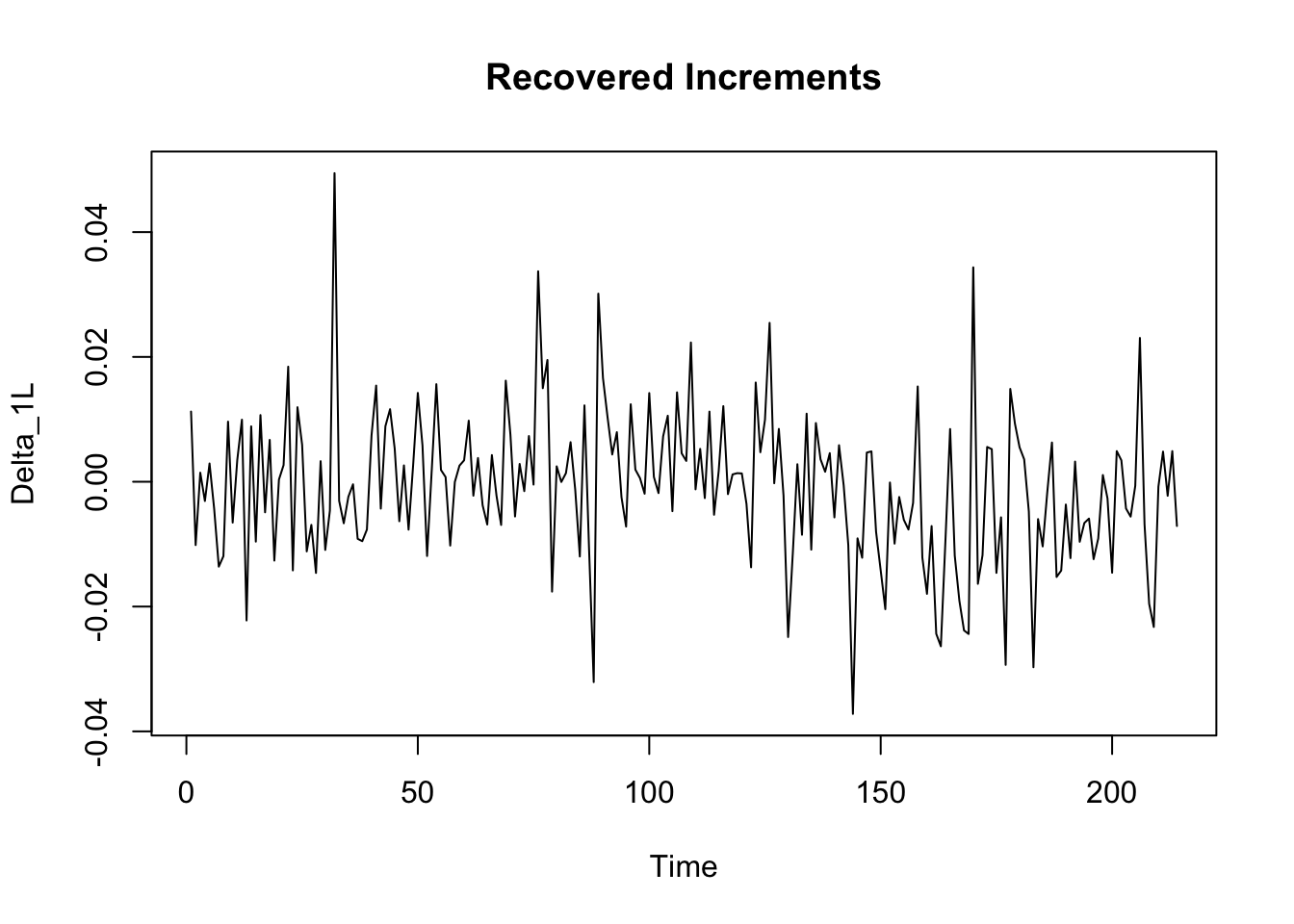}
    \caption{\label{ABTDHR} Recovered increments of ABT and DHR spread dynamic as a time series}
  \end{minipage}
  \hspace{0.05\textwidth} % Horizontal space between the two images
  \begin{minipage}[b]{0.45\textwidth}
    \centering
    \includegraphics[width=\textwidth]{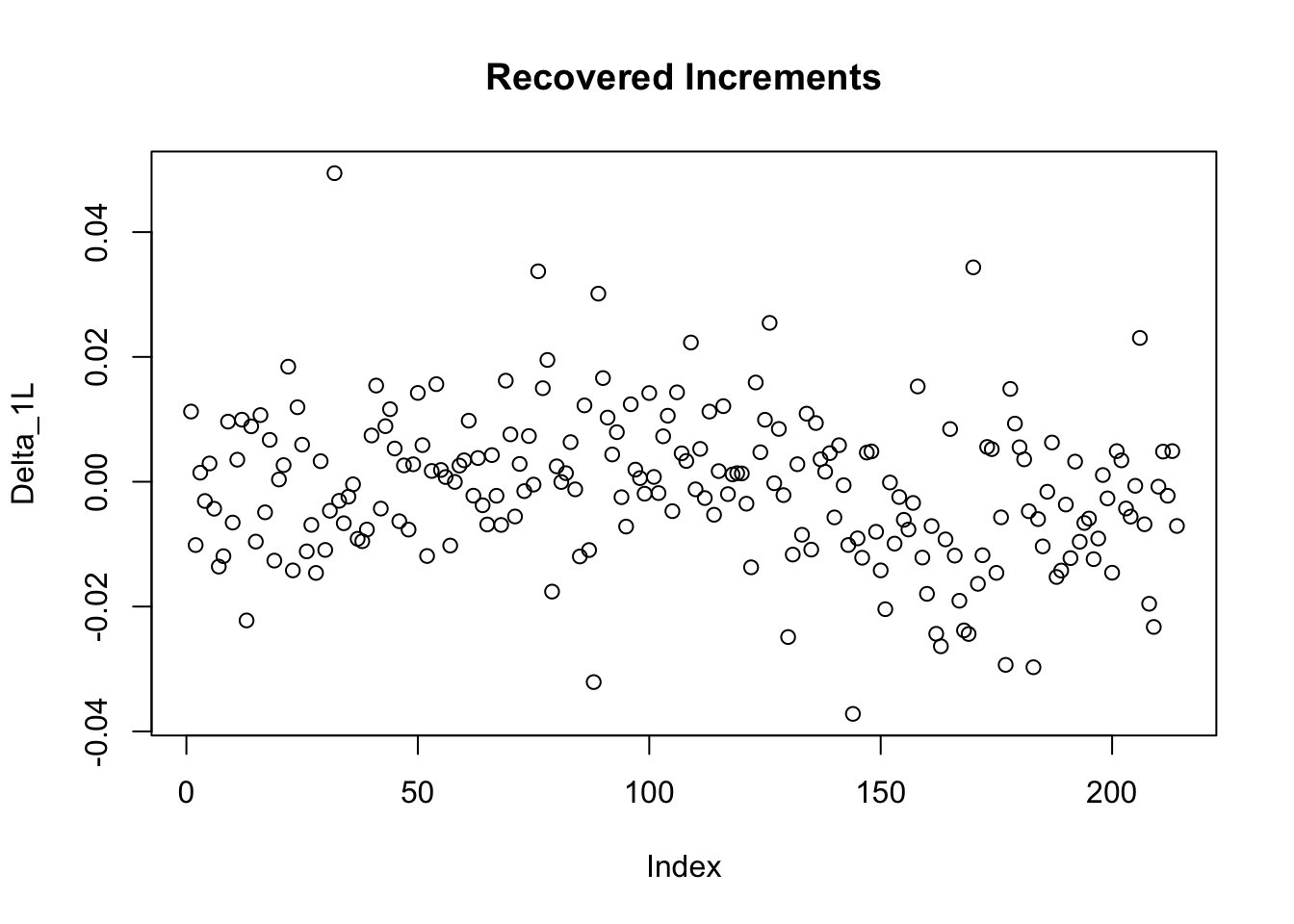}
    \caption{\label{Residuals1} Recovered increments of ABT and DHR spread dynamic as points}
  \end{minipage}
\end{figure}

We observe that the recovered increments of the ABT-DHR spread dynamics exhibit relatively random, uncorrelated behavior, suggesting that the L\'evy-driven CAR(1) model is a good fit. When plotted as points, the residuals of the recovered increments show no significant trends, outliers, or unusual clustering, further supporting the hypothesis of uncorrelated increments for this spread dynamic.\\

In Figure 5 below, we plot the sample autocorrelation function (ACF) for the recovered ABT-DHR spread dynamics.

\begin{figure}[H]
  \centering
  \includegraphics[width=0.64\textwidth]{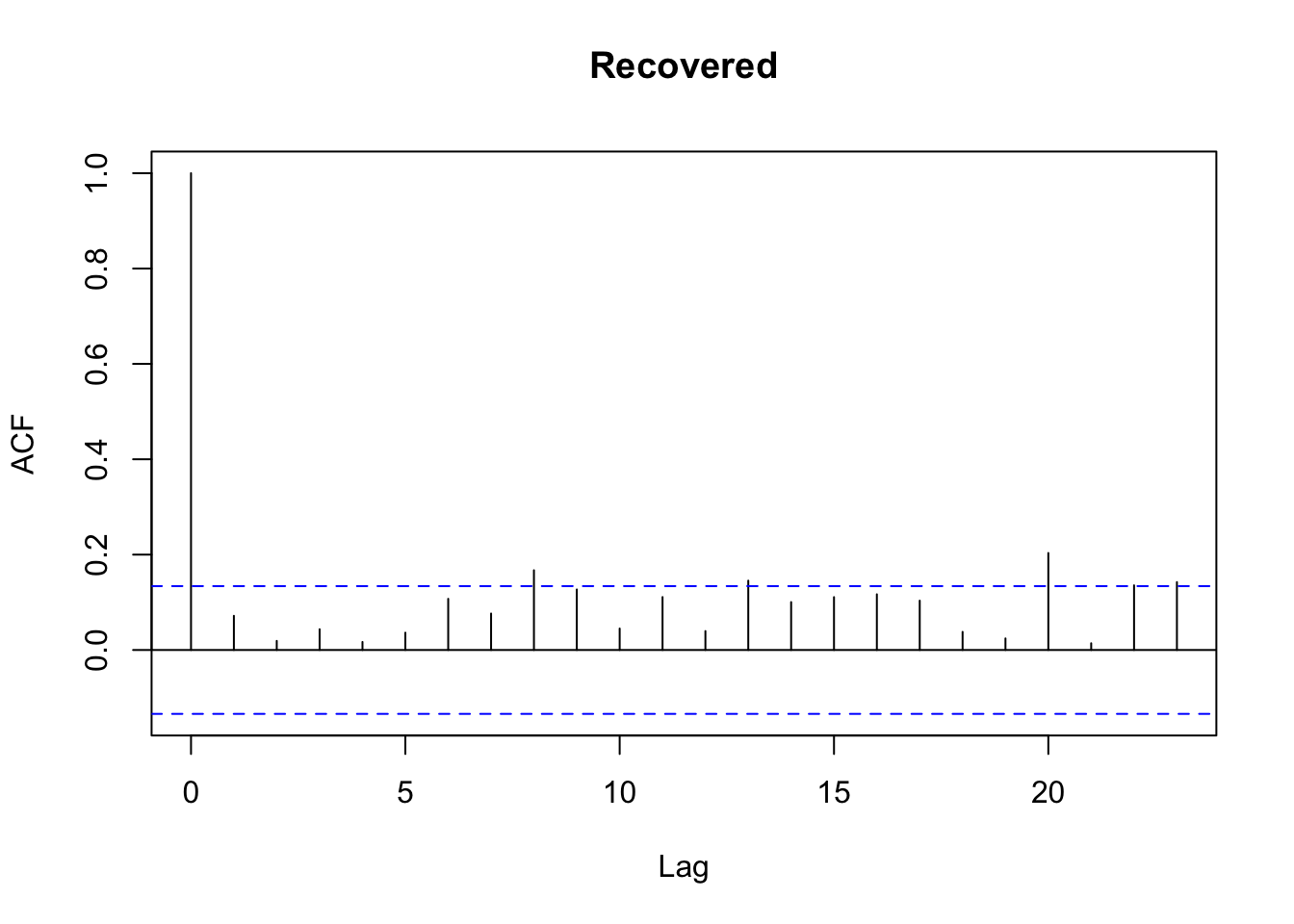}
  \caption{\label{ABTDHR2} The sample autocorrelation function for the recovered increments of ABT and DHR spread dynamic}
\end{figure}

With a test statistic of 1.05, well below the critical value of  $z = 1.96$, we fail to reject the null hypothesis, suggesting that a L\'evy-driven CAR(1) model could be a good fit for this spread dynamic. This is further supported by the autocorrelation function, which remains well below the dashed line at lag $=$ 1, allowing us to proceed to Step 5.\\

Following Step 5 (see Subsection  \ref{step 5}), we applied our test to assess whether the driving process of these spreads was Brownian motion. Using Procedure 1, we found that the spread dynamic of ABT and DHR failed to reject the normality assumption of the recovered increments, suggesting that the increments are consistent with a Brownian motion-driven process. We observed consistent results when applying Steps 1-5 to the spread of Apple (AAPL) and Google (GOOG). Additionally, for further confirmation, when we applied Procedure 2 to the spread of ABT and DHR to test for Brownian motion, the spread dynamic still failed to reject the normality assumption. Our applications of Procedures 1 and 2 to the data can be found in Sections 3.8.1 and 3.8.2 of the supplemental material.\\

On the other hand, the spread between Goldman Sachs Group (GS) and Morgan Stanley (MS) rejects the null hypothesis, producing a test statistic of 4.30. Similarly, the spread between Amgen Inc. (AMGN) and Pfizer (PFE) also rejects the null hypothesis with a test statistic of 4.78. This suggests that a  L\'evy-driven OU model may not be a good fit for these spreads. In Figures 6 and 7 below, we plot the recovered increments for the spread between Amgen and Pfizer, both as a time series and as residuals, respectively.

\begin{figure}[H]
  \centering
  \begin{minipage}[b]{0.45\textwidth}
    \centering
    \includegraphics[width=\textwidth]{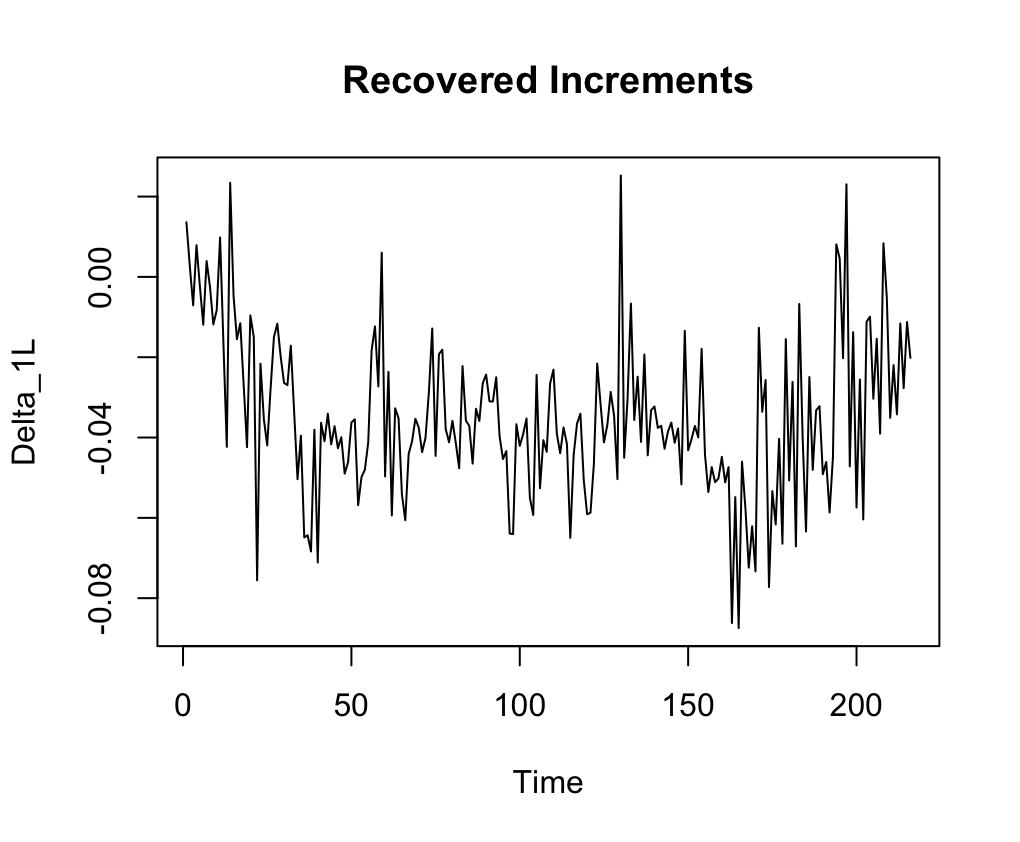}
    \caption{\label{AAAPGOOG} Recovered increments of Amgen and Pfizer spread dynamic as a time series}
  \end{minipage}
  \hspace{0.05\textwidth} % Horizontal space between the two images
  \begin{minipage}[b]{0.45\textwidth}
    \centering
    \includegraphics[width=\textwidth]{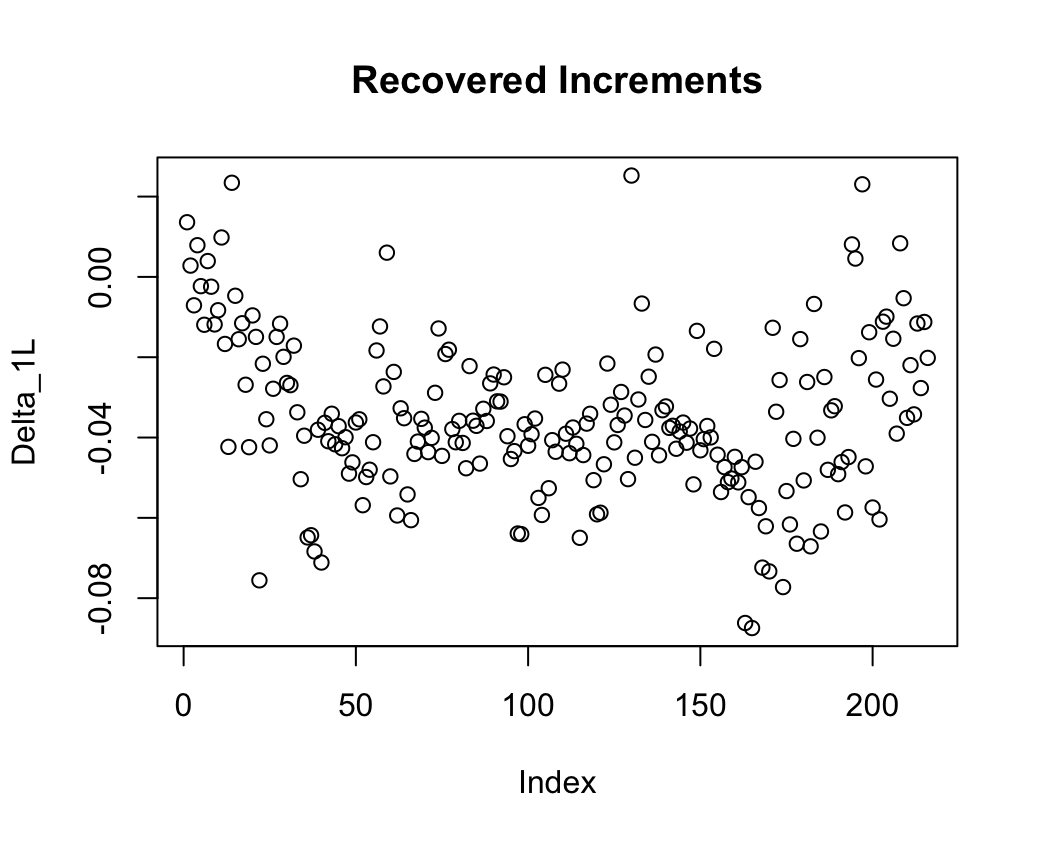}
    \caption{\label{Residuals1} Recovered increments of Amgen and Pfizer spread dynamic as points}
  \end{minipage}
\end{figure}

We observe a very clear leftover relationship in the recovered increments for the Amgen and Pfizer spread dynamic, suggesting that they are correlated. This observation is further supported by the autocorrelation function in Figure 8. The residuals plot shows clear trends and clustering, indicating potential correlations between increments that do not align with the uncorrelated assumption of a L\'evy-driven CAR(1) process.\\

%In Figure 8 below, we plot the sample autocorrelation function for the recovered spread dynamic of Amgen and Pfizer.

\begin{figure}[H] 
  \centering
  \includegraphics[width=0.60\textwidth]{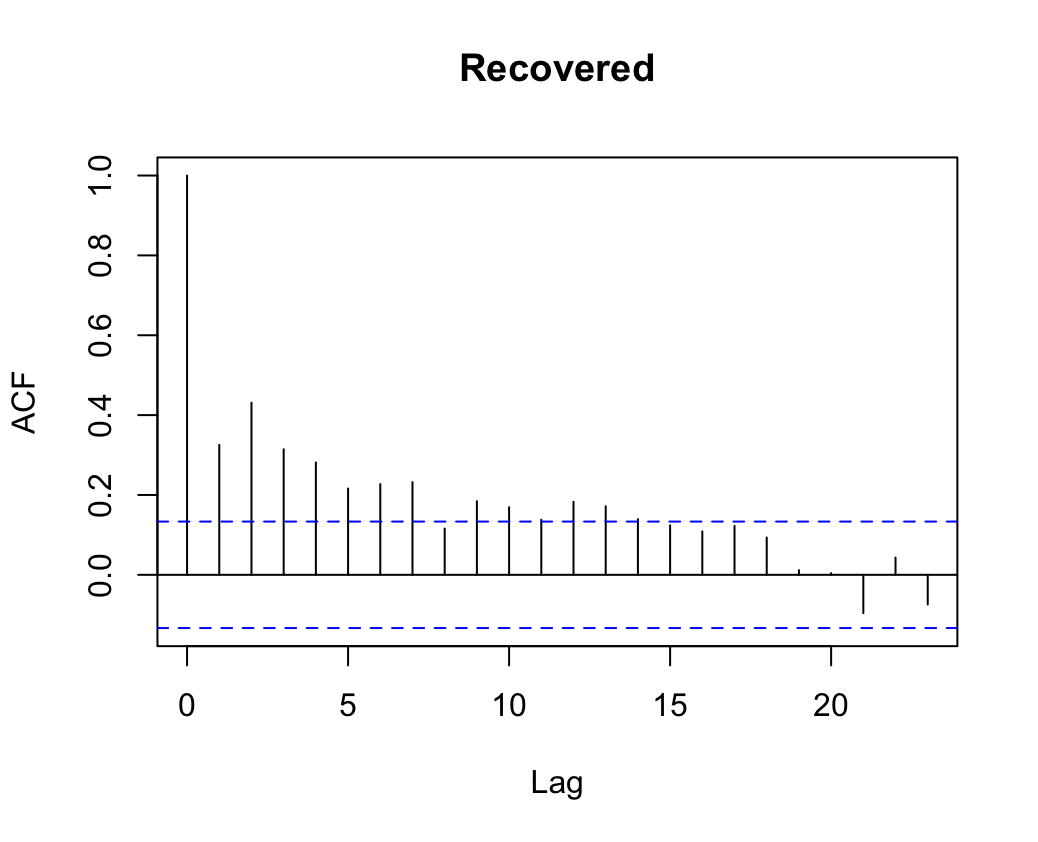}
  \caption{\label{AMGNPFE2} The sample autocorrelation function for the recovered increments of Amgen and Pfizer spread dynamic}
\end{figure}

With a test statistic of 4.78, which is well above the critical value of $z = 1.96$, we reject the null hypothesis, suggesting that a L\'evy-driven CAR(1) model may not be a good fit for this spread dynamic. This conclusion is further supported by the autocorrelation function, which is well above the dashed line at lag $=$ 1. Therefore, we do not need to proceed to Step 5.\\

{\bf Conclusion:}\\
The results vary significantly by GICS sector, but the verification process indicates that the L\'evy-driven CAR(1) model is not be a good fit for all spread processes. This includes certain individual pair combinations as well as some GICS sectors. However, many individual pair combinations and several GICS sectors perform well under the verification process, supporting the selection of a L\'evy-driven model for their spread processes. For further details, including R code for spread calculation and the application of the verification methodology to our data, refer to Section 3 of the supplemental material.\\

\section{Application to Realized Volatility of Euro/USD Exchange Rate}\label{sec:section 5}

\subsection{Data and model}

We next apply our verification process to the realized volatility calculated from Euro/USD currency exchange rate data. This data, sourced from Olsen Data, spans 10 years (from June 2007 to June 2017) and consists of Euro/USD bid and ask prices recorded at 5-minute intervals. In the Foreign Exchange (FX) market, the bid price represents the amount a currency dealer is willing to pay for a currency, while the ask price is the rate at which the dealer is willing to sell it.\\

After cleaning the data to remove missing entries, weekends, fixed holidays, and other calendar effects, we are left with 2,463 full trading days of 5-minute interval data, totaling approximately 709,555 data points.\\

With our cleaned dataset, we calculate daily returns, determine the corresponding realized variance, and then compute the square root to obtain the daily realized stochastic volatility.\\

Realized volatility is represented by the following equation:

\begin{equation}\label{volatility}
    RV = \sqrt{\sum_{i=1}^{N}r_t^2},
\end{equation}

where $r_t$ is the daily return at time. In this context, realized volatility measures the variation in returns for the Euro/USD currency exchange rate. Below, in Figure 9, we model the Euro/USD returns for each 5-minute interval, calculated from our data:

\begin{figure}[H]
  \centering
  \includegraphics[width=0.64\textwidth]{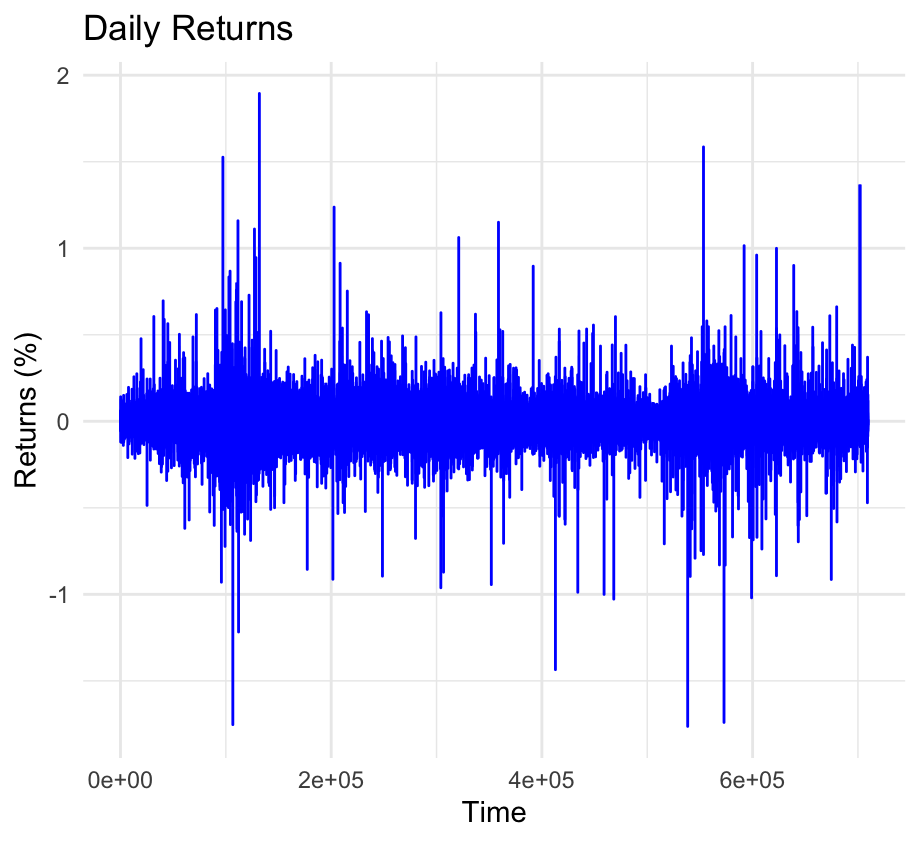}
  \caption{\label{Spread1} Euro/USD 5-minute returns}
\end{figure}

We then follow Equation (\ref{volatility})  to calculate realized volatility. The daily realized volatility is modeled below in Figure 10:

\begin{figure}[H]
  \centering
  \includegraphics[width=0.64\textwidth]{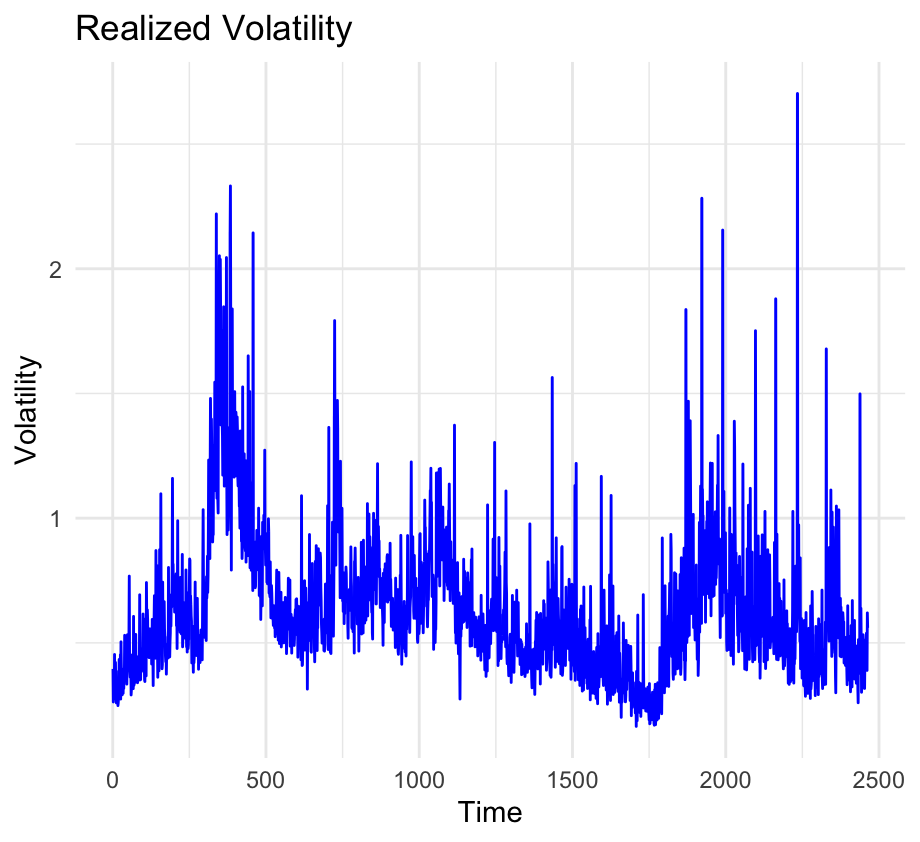}
  \caption{\label{Spread2} Euro/USD daily realized volatility}
\end{figure}

\subsection{Verification}

We then follow the methodology outlined in Section 3 to verify whether the realized volatility follows a L\'evy-driven CAR(1) process. In Step 1, we choose the DMB estimator for 
$a$, as realized volatility only takes on positive values and exhibits characteristics consistent with Gamma-driven and Inverse Gaussian-driven processes. In Step 2, we set the number of large periods to $N=35$ and the sampling frequency to $M=70$, where $\frac{2463}{M}=N$, with 2463 representing the total number of days in our cleaned dataset. In Step 3, we recover the increments for the Euro/USD realized volatility using the chosen values and estimated parameters. Finally, in Step 4, we calculate the test statistic and test the correlation of the recovered increments.\\

In Figures 11 and 12 below, we model the recovered increments of the Euro/USD realized volatility both as a time series and as residuals.

\begin{figure}[H]
  \centering
  \begin{minipage}[b]{0.45\textwidth}
    \centering
    \includegraphics[width=\textwidth]{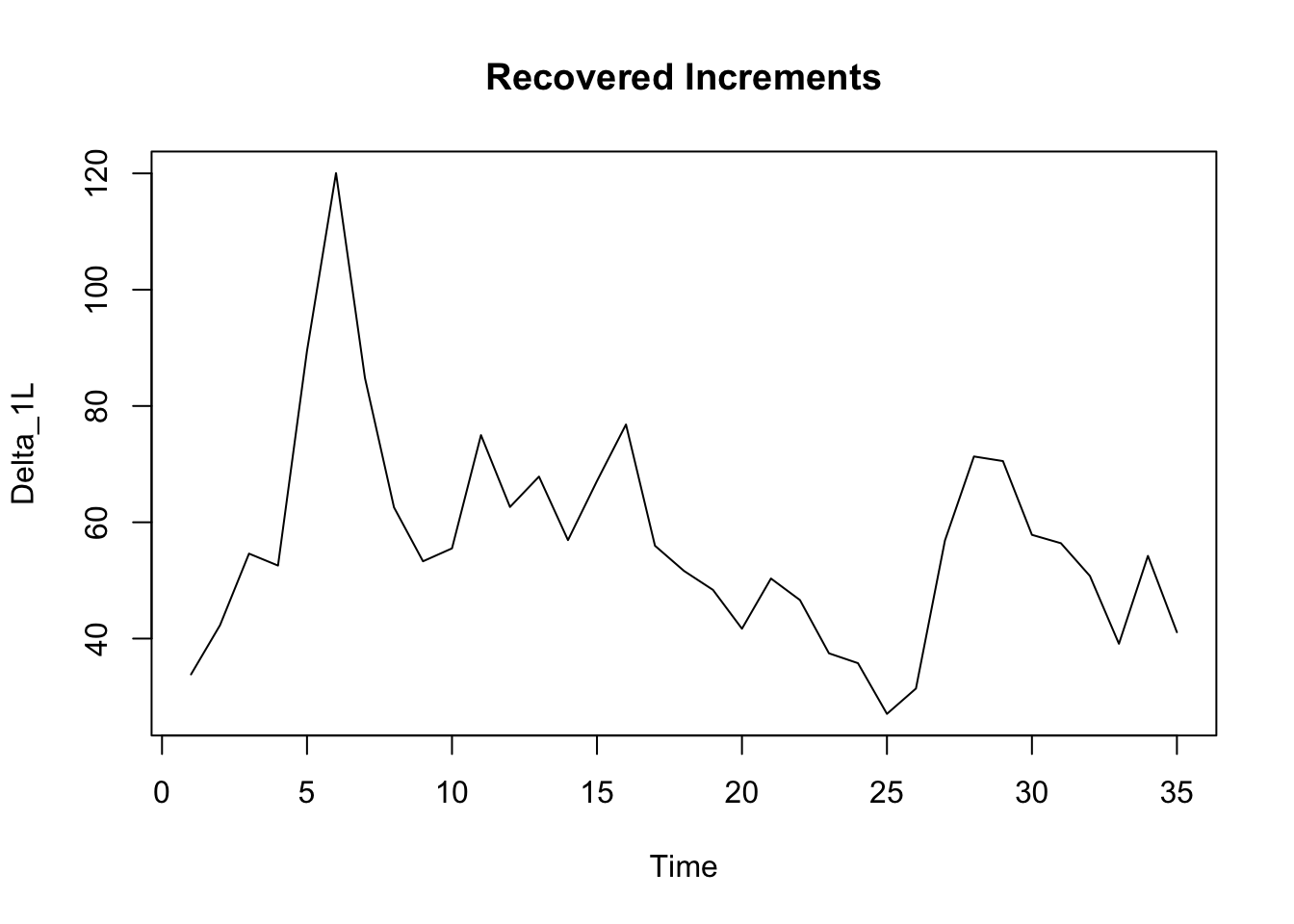}
    \caption{\label{EUROUSD} Recovered increments of Euro/USD realized volatility as a time series}
  \end{minipage}
  \hspace{0.05\textwidth} % Horizontal space between the two images
  \begin{minipage}[b]{0.45\textwidth}
    \centering
    \includegraphics[width=\textwidth]{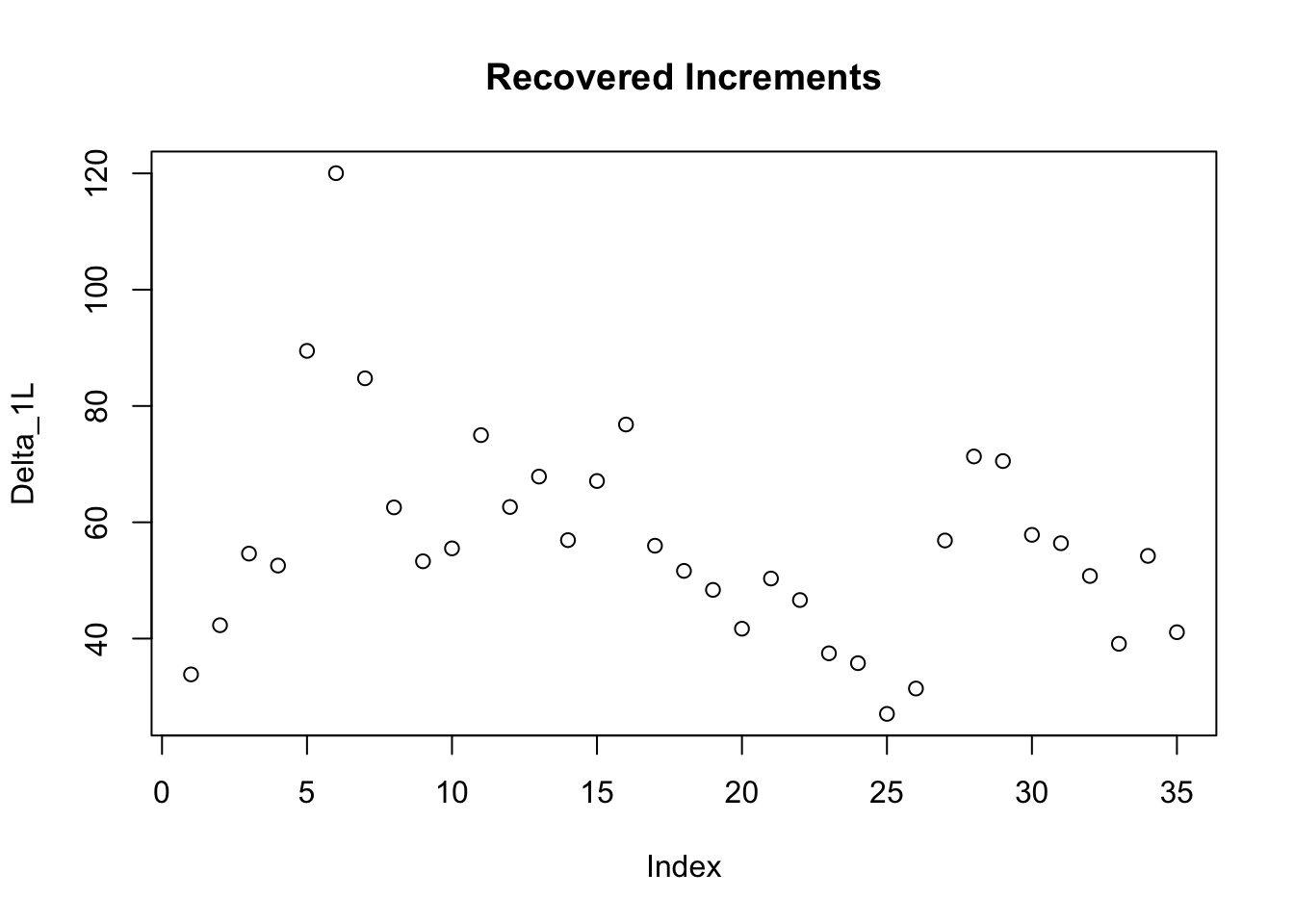}
    \caption{\label{Recovered1} Recovered increments of Euro/USD realized volatility as points}
  \end{minipage}
\end{figure}

We observe that the recovered increments of the Euro/USD realized volatility exhibit noticeable correlation over time, suggesting that the increments are not independent. When plotted as individual points, the residuals show a distinct pattern, indicating that the increments may exhibit correlation. This suggests that a L\'evy-driven CAR(1) model may not be a good fit, a conclusion supported by the autocorrelation function in Figure 13.\\
%In Figure 13 below, we plot the sample autocorrelation function for the recovered Euro/USD realized volatility. 

With a test statistic of 3.76 for the Euro/USD realized volatility, which is well above the critical value $z = 1.96$, we reject the null hypothesis. This suggests that a L\'evy--driven CAR(1) model may not be a good fit for this spread dynamic. This is confirmed by the autocorrelation function being well above the dashed line at lag $=$ 1, and thus, we do not need to proceed to Step 5.\\

\begin{figure}[H] 
  \centering
  \includegraphics[width=0.64\textwidth]{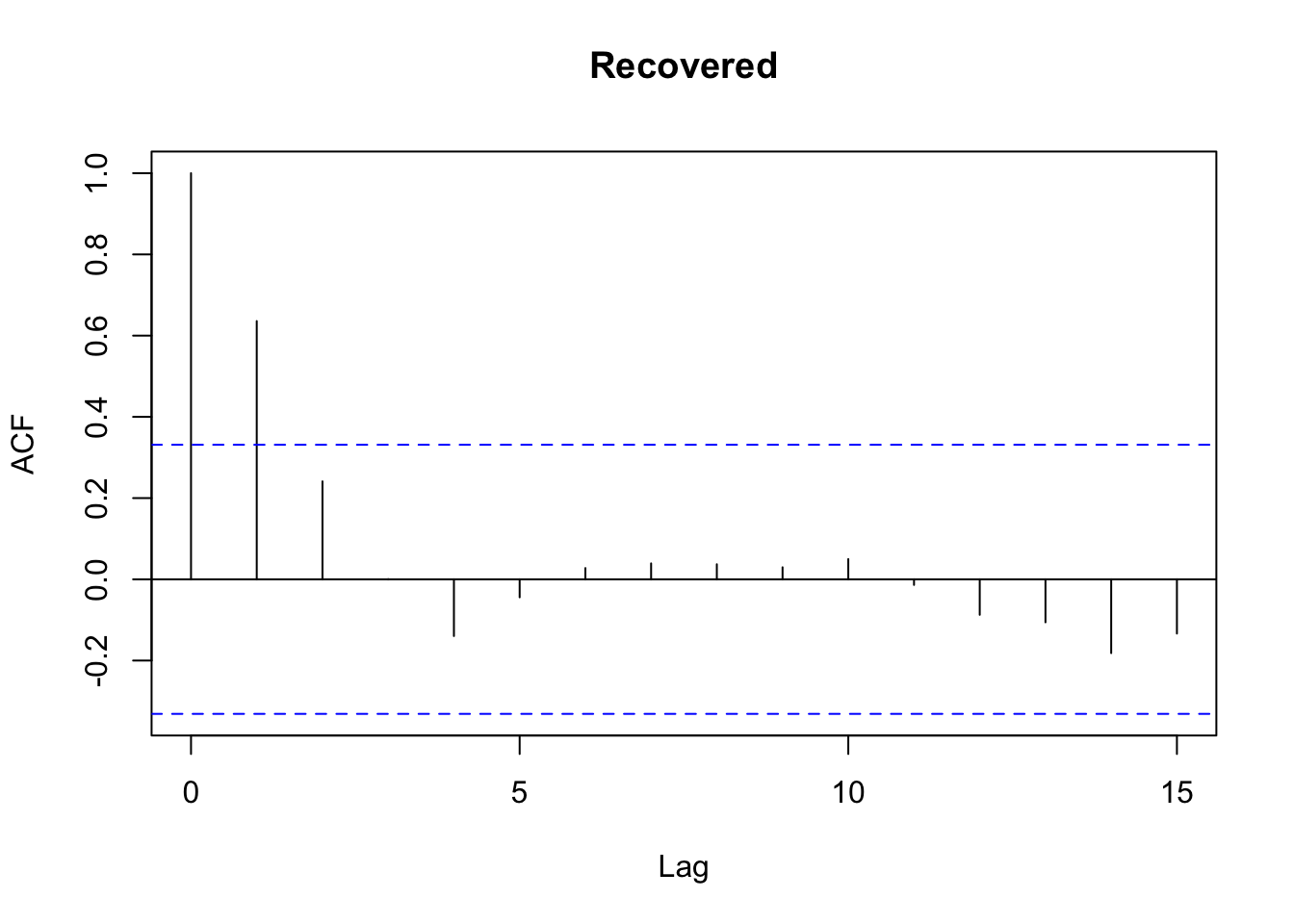}
  \caption{\label{ExchangeACF} The sample autocorrelation function for the recovered increments of Euro/USD realized volatility}
\end{figure}

In line with \cite{Brockwelletall2007}, we believe that the L\'evy-driven CARMA(2,1) process could be a good fit for realized volatility. However, further verification is needed to confirm that the Lévy-driven CARMA(2,1) model is indeed a suitable fit, and this work will be conducted in the near future.\\

\section{Conclusion}

In conclusion, our verification process proves highly effective in testing the fit of models assumed to be L\'evy-driven. In this paper, we apply the process to test a L\'evy-driven model for the price spread between S\&P 500 stock pairs and Euro/USD exchange rate data. However, the verification process can be extended to other real-world data, such as interest rates, credit risk, population dynamics, and more.\\

Our tests yield strong results, failing to reject the null hypothesis of uncorrelated increments in most cases, approximately 95\% of the time at the nominal level ($\alpha=0.05$), across many simulated L\'evy-driven CAR(1) background processes, including Brownian motion, Gamma, Beta, Inverse Gaussian, and mixed combinations of these processes. Additionally, when testing for Brownian motion, Procedures 1 and 2 produced p-values approaching 0.05 under $H_0$. Moreover, when testing the power of these two procedures, they rejected the normal distribution hypothesis at high rejection rates when the recovered increments deviated from a non-normal distribution for varying combinations of $N$, $M$, and $a$.\\

Our methodology proves to be a powerful tool for verifying whether a given process is an Ornstein-Uhlenbeck process, with direct application to real-world data. This paper provides the precise methodology for applying this test, and our supplementary material includes further explanations as well as R code corresponding to the verification process and the data examples.\\

In future work, we plan to write a user-accessible methodological paper detailing how to verify that a given process is a L\'evy-driven CARMA(2,1) process.\\

\bibliographystyle{plain}
\bibliography{References-3}{}

\begin{thebibliography}{10}

\bibitem{Ibrahim}
I.~Abdelrazeq.
\newblock Model verification for {L}\'evy-driven {O}rnstein-{U}hlenbeck
  processes with estimated parameters.
\newblock {\em Statistics and Probability Letters}, 104:26--35, 2015.

\bibitem{AIK}
I.~Abdelrazeq, B.G. Ivanoff, and R.~Kulik.
\newblock Model verification for {L}\'evy-driven {O}rnstein-{U}hlenbeck
  processes.
\newblock {\em Electronic Journal of Statistics}, 8:1029--1062, 2014.

\bibitem{CJS2018}
Ibrahim Abdelrazeq, B.~Gail Ivanoff, and Rafal Kulik.
\newblock Goodness-of-fit tests for {L}\'{e}vy-driven {O}rnstein-{U}hlenbeck
  processes.
\newblock {\em Canad. J. Statist.}, 46(2):355--376, 2018.

\bibitem{Barndoroffandshephard2001}
Ole~E. Barndorff-Nielsen and Neil Shephard.
\newblock Non-{G}aussian {O}rnstein-{U}hlenbeck-based models and some of their
  uses in financial economics.
\newblock {\em J. R. Stat. Soc. Ser. B Stat. Methodol.}, 63(2):167--241, 2001.

\bibitem{Brockwelletall2007}
Peter~J. Brockwell, Richard~A. Davis, and Yu~Yang.
\newblock Estimation for nonnegative {L}\'evy-driven {O}rnstein-{U}hlenbeck
  processes.
\newblock {\em J. Appl. Probab.}, 44(4):977--989, 2007.

\bibitem{BurkeGombay1998}
Murray~D. Burke and Edit Gombay.
\newblock On goodness-of-fit and the bootstrap.
\newblock {\em Statist. Probab. Lett.}, 6(5):287--293, 1988.

\bibitem{DavisMcCormick}
R.~A. Davis and W.~P. McCormick.
\newblock Estimation for first-order autoregressive processes with positive or
  bounded innovations.
\newblock {\em Stochastic Process, Appl.}, 31(2):237--250, April 1989.

\bibitem{Khmaladze2021}
E.~V. Khmaladze.
\newblock How to test that a given process is an ornstein–uhlenbeck process.
\newblock {\em Stat Inference Stoch Process}, 24(3):405--419, July 2021.

\bibitem{QuantQuote2016}
QuantQuote.
\newblock Quantquote market data and software.
\newblock \url{https://www.quantquote.com/}, 2016.

\bibitem{Stuteetal}
W.~Stute, W.~Gonz\'{a}les Manteiga, and M.~Presedo Quindimil.
\newblock Bootstrap based goodness-of-fit tests.
\newblock {\em Metrika}, 40(3-4):243--256, 1993.

\end{thebibliography}

\end{document}